\documentclass[aip,jcp,amsmath,amssymb,floatfix,preprint]{revtex4-2}

\usepackage{graphicx}% Include figure files
\usepackage{dcolumn}% Align table columns on decimal point
\usepackage{bm}% bold math
\usepackage{color}
\def\be{\begin{equation}}
\def\en{\end{equation}}
\renewcommand{\theequation}{\arabic{section}.\arabic{equation}}
\def\p{\partial}
\def\ep{\epsilon}

\def\ve{\varepsilon}
\newcommand{\av}[1]{\langle{#1}\rangle}

\newcommand{\pp}[2]{\frac{\partial {#1}}{\partial {#2}}}
\newcommand{\ppm}[2]{\frac{\delta{#1}}{\delta{#2}}}

\usepackage[utf8]{inputenc}
\usepackage[T1]{fontenc}
\usepackage{mathptmx}
\usepackage{etoolbox}
\makeatletter
\@addtoreset{equation}{section}
\makeatother

\begin{document}

\title{Enhancement of liquid-liquid phase transition under shear flow}
\author{Kyohei Takae}
\email{takae@tottori-u.ac.jp}
\affiliation{Graduate School of Engineering, Tottori University, 4-101 Koyama-Cho Minami, Tottori, Tottori 680-8552, Japan}
\affiliation{Advanced Mechanical and Electronic System Research Center, Faculty of Engineering, Tottori University, 4-101 Koyama-Cho Minami, Tottori, Tottori 680-8552, Japan}
\date{\today}

\begin{abstract}
In a liquid--liquid transition (LLT), a single-component liquid transforms into another liquid with different density and viscosity. Flow therefore couples to the transition, but the effect of shear flow is not understood. Here we investigate the LLT under shear with a two-order-parameter hydrodynamic model in which the viscosity depends on the non-conserved structural order parameter. Under shear, a fluctuation of the order parameter drives a compressional flow through the viscosity, and the resulting density change enhances the order parameter fluctuation. In a linear analysis, this feedback destabilizes the metastable liquid above a critical shear rate $\dot{\gamma}_{\rm c}$ for wavevectors in particular directions. Because the flow rotates the wavevector, a fluctuation grows only for a finite time duration, and its amplification is determined by the dimensionless shear rate $\dot{\gamma}/\dot{\gamma}_{\rm c}$ and the second dimensionless number $\Lambda$, the ratio of the thermodynamic relaxation rate of the order parameter to $\dot{\gamma}_{\rm c}$. In hydrodynamic simulations, shear above about $\dot{\gamma}_{\rm c}$ induces the transition in the metastable region when $\Lambda\gtrsim1$. The transition proceeds by nucleation of domains whose morphology becomes anisotropic. The incubation time decreases and the onset of the viscosity increase is accelerated as $\dot{\gamma}$ and $\Lambda$ increase. These results imply that the shear rate is a control variable of LLT in addition to temperature and pressure, selecting whether and how a metastable liquid transforms.
%is determined by the linear amplification and the kinetic coefficient. For smaller $\Lambda$, the amplification is weak, and the liquid remains metastable with anisotropically enhanced fluctuations. Since the absolute viscosity cancels from $\Lambda$, the condition $\Lambda\gtrsim1$ can be satisfied in viscous liquids such as molecular liquids and metallic melts near their glass transition.
\end{abstract}

\maketitle

\section{Introduction}
The liquid state of matter has long been viewed as structurally featureless. A single-component liquid was assumed to exist as one phase, whose properties vary continuously with temperature and pressure. The discovery of liquid--liquid transitions (LLTs)---phase transitions between two distinct liquid states of a single-component substance---has changed this view~\cite{tanaka2020liquid}. The best-known example is water~\cite{gallo2016water}, for which a transition between low- and high-density liquids and an associated second critical point were hypothesized~\cite{poole1992phase,mishima1998decompression}. Although the transition itself lies in a region hindered by crystallization, this hypothesis has recently been supported experimentally by ultrafast X-ray probing of bulk supercooled water~\cite{kim2020experimental,amann-winkel2023liquid} and by the observation of critical fluctuations and a diverging heat capacity near the predicted critical point~\cite{you2026experimental}. LLTs have also been observed or predicted in atomic liquids such as phosphorus~\cite{katayama2000first}, sulfur~\cite{henry2020liquid}, silicon~\cite{sastry2003liquid}, and carbon~\cite{glosli1999liquid}, in molecular liquids such as triphenyl phosphite (TPP)~\cite{cohen1996low,tanaka2004liquid,kobayashi2016reversibility,walton2020polyamorphism}, and in metallic glass-forming liquids~\cite{shen2020metallic,dong2021situ}. The LLT is thus a general phenomenon of condensed matter. Because the two liquids differ in density, structure, and mobility, the transition between them affects glass formation and crystallization. Therefore, the kinetics of the LLT, that is, how one liquid transforms into the other, is essential for controlling materials properties within the liquid state.

Theoretically, LLTs are commonly described by two-state, or two-order-parameter, models~\cite{tanaka2000general,shi2020anomalies,anisimov2018thermodynamics,caupin2021minimal,li2026evidence}. In these models a liquid is characterized by its density and by the fraction of locally favored structures. The latter is a non-conserved order parameter and undergoes the transition. The density change at the LLT follows from the coupling between the order parameter and the density. These models explain the thermodynamic anomalies of water and other polymorphic liquids. The existence of the transition and of its critical point has been established in water models by molecular dynamics simulations~\cite{debenedetti2020second,gartner2022liquid,sciortino2025constraints,wang2026direct}. 
Experimental observations of nucleation--growth and spinodal-decomposition regimes, and their crossover with quench depth, in molecular liquids such as TPP are consistent with this description~\cite{kurita2004critical,kurita2006kinetics,kurita2007phase,kobayashi2016reversibility}.
However, these theories, including the recent treatments of phase-ordering kinetics in inter-converting two-state systems~\cite{shumovskyi2021phase}, have been formulated for quiescent liquids, with the exception of the hydrodynamic theory of Refs.~\cite{takae2011phase,takae2020role}. Because the two liquids differ in density, the transition is accompanied by fluid flow, which alters the domain growth kinetics, as in the crystallization of a liquid~\cite{oxtoby1992effect}. Hydrodynamic flow is thus intrinsic to LLT, yet the effects of externally imposed flow on the LLT still remain elusive.

Shear flow is known to modify phase transitions in fluids, in a way that depends on the system~\cite{Onukibook}. In near-critical fluids and mixtures, shear generally suppresses long-wavelength fluctuations and distorts phase separation into anisotropic patterns~\cite{onuki1979nonequilibrium,Onukibook}. By contrast, in dynamically asymmetric systems, where the viscosity depends strongly on a conserved variable such as the composition or the density, shear enhances fluctuations and destabilizes the homogeneous state~\cite{Onukibook,onuki1989elastic,helfand1989large,milner1993dynamical,furukawa2006violation,furukawa2009inhomogeneous}. For crystallization in atomic liquids and colloidal systems, shear is known to either accelerate or suppress nucleation depending on the shear rate, through its competing effects on the thermodynamic driving force and on transport coefficients~\cite{mokshin2010crystal,goswami2021homogeneous,bhendale2025shear,holmqvist2005crystallization,blaak2004crystal,richard2015role,shao2015shear}. A transient growth of fluctuations under shear is also known for the gravitational instability of galactic gas disks~\cite{kim2001amplification}. Shear also affects the polyamorphism of water: the medium-density amorphous solid produced by ball-milling of ice~\cite{rosu-finsen2023medium} has been identified in molecular simulations as a nonequilibrium steady state whose density is governed by the shear rate~\cite{ribeiro2024medium}. However, these molecular simulations were performed deep in the glassy state, and how shear affects the LLT in the liquid state has not been investigated. Experimentally, LLTs have been accessed by changing temperature or pressure, and the influence of flow on a proposed LLT has been examined only in the bulk metallic glass former Zr$_{41.2}$Ti$_{13.8}$Cu$_{12.5}$Ni$_{10}$Be$_{22.5}$~\cite{way2007influence,gallino2021effect}. In this system, steady shear shortens the incubation time of the transition from the fragile, low-viscosity liquid to the strong liquid, and the acceleration was described with the classical nucleation theory extended to sheared systems~\cite{mura2016effects}. In that theory, shear enters through the transport of the molecules and the displacement of the nucleus, whereas the density coupling and the viscosity change are not considered. In the LLT, the viscosity depends on the structural order parameter, which is non-conserved and is hydrodynamically coupled to the density. How shear flow affects the transition through these couplings has not been understood.

In this paper, we investigate the effect of shear flow on LLTs based on the two-order-parameter continuum model, incorporating the dependence of the viscosity on the order parameter~\cite{takae2020role}. In this model, a fluctuation of the structural order parameter induces an inhomogeneous viscous stress under shear. The stress drives a compressional flow, and the resulting density change enhances the order parameter fluctuation. Above a critical shear rate, this feedback exceeds the thermodynamic restoring force for wavevectors in a particular direction. Because the flow rotates the wavevector of the unstable modes, the feedback acts on a fluctuation for a limited time duration, and the growth during this time depends on the timescale of the order parameter relaxation. A linear theory and nonlinear hydrodynamic simulations show that the effect of shear is governed by two numbers, the shear rate relative to the critical shear rate and the ratio of the order parameter relaxation rate to the critical shear rate. When both numbers are of order unity or larger, shear promotes the nucleation in the metastable liquid. When the relaxation is slower, the liquid remains metastable, but its fluctuations are enhanced anisotropically. The shear rate thus acts as a control variable of the LLT in addition to temperature and pressure.

\section{Theory}
\label{sec:theory}
\subsection{Ginzburg--Landau free energy}
\label{sec:GL}
Our description of LLT relies on the two-order-parameter picture of a liquid~\cite{kurita2007phase,takae2020role}, in which the state is specified by the mass density $\rho$ (conserved) together with the fraction $c$ of locally favored structures (non-conserved). The hydrodynamic formulation of Ref.~\cite{takae2020role} is adopted in its isothermal form and is summarized below.

The Landau expansion is taken at the bare spinodal point of the structural order parameter without density coupling, $(T_0,\rho_0)$, where $c=c_0$. In terms of $\phi=c-c_0$ and $m=\rho/\rho_0-1$, the dimensionless free-energy density reads
\begin{align}
\frac{v_0}{k_{\rm B}T_0}f=&f_0(\phi)-\alpha\phi m +\mu_0 n_0 m +\frac{1}{2K_0}m^2,\label{eq:f}\\
f_0(\phi)=&h\phi+\frac{\kappa}{2}\phi^2+\frac{b_3}{3}\phi^3+\frac{b_4}{4}\phi^4,\label{eq:f0}
\end{align}
where $v_0$ is the molecular volume and $k_{\rm B}$ is the Boltzmann constant. The polynomial $f_0$ is the expansion of the two-state free energy~\cite{kurita2007phase} at $c_0$, which gives $h=[\Delta s/k_{\rm B}+\ln(c_0/(1-c_0))]\tau$, $\kappa=\tau/[c_0(1-c_0)]$, $b_3=-[1/c_0^2-1/(1-c_0)^2]/2$, and $b_4=[1/c_0^3+1/(1-c_0)^3]/3$. Here $\Delta s$ is the entropy difference between liquid 1 and liquid 2, and $\tau=T/T_0-1$ is the reduced temperature. The values $\Delta s/k_{\rm B}=4$ and $c_0=0.2$ are used, so that $h=2.61\tau$, $\kappa=6.25\tau$, $b_3=-11.7$, and $b_4=42.3$~\cite{kurita2007phase}. The bilinear term $-\alpha\phi m$ denotes the coupling between the order parameter and the density, giving the density difference between the two liquids: liquid 2 has a higher (lower) density than liquid 1 for $\alpha>0$ ($\alpha<0$). $K_0$ is the dimensionless isothermal compressibility of the reference state, while $\mu_0$ and $n_0$ (reference chemical potential and number density) add only a constant to the pressure and therefore drop out of the equations of motion. Due to the $\phi$--$m$ coupling, the spinodal temperature $\tau_{\rm s}$ is displaced from its bare location $\tau=0$. The free-energy functional in its isothermal form reads
\be
\beta \mathcal{F}=\frac{1}{\ell_0^d}\int d^d\bm r \left[\beta v_0 f(\phi,m)+\frac{D}{2}|\nabla\phi|^2\right],
\label{eq:F}
\en
where $\beta=1/k_{\rm B}T_0$ is the inverse temperature, $\ell_0$ is a characteristic length comparable to the correlation length, and the square-gradient term together with $f$ gives the interfacial tension and interfacial width between the two liquids.

\subsection{Hydrodynamic equations}
\label{sec:hydro}
Since $\rho$ takes different values in the two liquids, LLT is accompanied by compressible flow, and $\phi$ must be evolved together with the full compressible hydrodynamics. In the isothermal limit, the equations of motion reduce to~\cite{takae2020role}
\begin{align}
\pp{}{t}\rho&=-\nabla\cdot(\rho\bm v),\label{eq:cont}\\
\pp{}{t}(\rho \bm v)&=-\nabla\cdot(\rho\bm v\bm v +\tensor\Pi-\tensor\sigma-\tensor\sigma^{\rm R}),\label{eq:mom}\\
\pp{}{t}\phi&=-\bm v\cdot \nabla\phi-L\Gamma + \theta.\label{eq:TDGL}
\end{align}
Equation~\eqref{eq:cont} expresses mass conservation and Eq.~\eqref{eq:mom} momentum conservation, with the reversible stress $\tensor\Pi$ and the viscous stress $\tensor\sigma$ given by
\begin{align}
\beta v_0\Pi_{ij}=&\beta v_0\hat{p}\delta_{ij} + D (\nabla_i\phi)(\nabla_j\phi),\label{eq:Pi}\\
\sigma_{ij}=&\eta(\phi)(\nabla_iv_j+\nabla_jv_i) +[\zeta(\phi)-2\eta(\phi)/d](\nabla\cdot\bm v) \delta_{ij},\label{eq:sigma}
\end{align}
where $d$ is the spatial dimension. The isotropic part of $\tensor\Pi$, $\beta v_0\hat{p}=\beta v_0 p-D |\nabla\phi|^2/2$, consists of the local thermodynamic pressure
\be
\beta v_0 p=\beta v_0[n\pp{f}{n}-f]=-f_0(\phi)-\alpha\phi+\mu_0n_0+\frac{m}{K_0}+\frac{m^2}{2K_0},
\label{eq:pre}
\en
and the gradient contribution. The first and second terms denote the pressure response to a change in local structure. The shear viscosity $\eta(\phi)$ and bulk viscosity $\zeta(\phi)$ are assumed to depend on the order parameter. For simplicity, the density dependence of the viscosity, which triggers the mechanical instability under shear in compressible liquids~\cite{furukawa2006violation}, is not incorporated. The $\phi$ dependence describes the viscosity contrast between the two liquids and is the origin of the shear-induced instability as explained below. Equation~\eqref{eq:TDGL} is relaxational (model A) dynamics for the non-conserved order parameter $\phi$, advected by the flow and driven by
\be
\Gamma=\ppm{\beta\cal F}{\phi}=f_0'(\phi)-\alpha m -D\nabla^2\phi,
\label{eq:Gamma}
\en
with kinetic coefficient $L$. The kinetic coefficient is assumed to be small, $L\ll1$, so that the relaxation of $\phi$ (formation of locally favored structures) is the slowest process and the transition is limited by the non-conserved kinetics rather than by momentum or density relaxation. The random stress $\tensor\sigma^{\rm R}$ and the random force $\theta$ are Gaussian white noises satisfying the fluctuation--dissipation theorem~\cite{Landau6,Onukibook},
\begin{align}
\av{\theta(\bm r,t)\theta(\bm r',t')}&=2L\delta(\bm r-\bm r')\delta(t-t'),
\label{eq:fluc1}
\\
\av{\sigma^{\rm R}_{ij}(\bm r,t) \sigma^{\rm R}_{k\ell}(\bm r',t')}
&=2k_{\rm B}T[\eta(\delta_{ik}\delta_{j\ell}+\delta_{i\ell}\delta_{jk})\nonumber\\
&+(\zeta-2\eta/d)\delta_{ij}\delta_{k\ell}]\delta(\bm r-\bm r')\delta(t-t').
\label{eq:fluc2}
\end{align}
%The noise terms are omitted in the linear stability analysis (Secs.~\ref{sec:instability} and \ref{sec:advection}). The random force $\theta$ is retained in the calculation of the structure factor (Sec.~\ref{sec:sk-theory}), and both noise terms are incorporated in the simulations.

\subsection{Shear-induced instability}
\label{sec:instability}
We examine the shear-induced phase-transition instability using linearized hydrodynamic equations without the random stress and force. To do this, the reference state at which the equations are expanded needs to be specified. In the NVT ensemble, the space average of the mass density $\av{m}$ is fixed, which is denoted by $m_{\rm e}$. The instability condition derived below has the same form in the NPT ensemble, with the coefficients evaluated at the isobaric state. Without flow, the homogeneous state of liquid~1 is $\phi=\phi_{\rm e}$, the solution of $\p f/\p \phi=0$ at $m=m_{\rm e}$. Under shear flow, the velocity field is decomposed as $\bm v=\dot{\gamma}y\bm e_x +\bm V$, where $\dot{\gamma}y\bm e_x$ is the imposed shear flow with $\dot{\gamma}$ being the shear rate. The linearized hydrodynamic equations in terms of $\delta m=m-m_{\rm e}$, $\psi=\phi-\phi_{\rm e}$, and $\bm V$ read
\begin{align}
\bigg(\pp{}{t}+\dot{\gamma}y\pp{}{x}\bigg)\delta m &= -(1+m_{\rm e})Z,\label{eq:lm}\\
\bigg(\pp{}{t}+\dot{\gamma}y\pp{}{x}\bigg)\psi &= -L[\kappa'-D\nabla^2]\psi +L\alpha \delta m,\label{eq:lpsi}\\
\rho_{\rm e}\bigg(\pp{}{t}+\dot{\gamma}y\pp{}{x}\bigg)\bm V &= -\rho_{\rm e}\dot{\gamma}V_y\bm e_x
+(1+m_{\rm e})\nabla[\alpha\psi-\delta m/K_0]\nonumber\\
&+\dot{\gamma}(\bm e_y\nabla_x+\bm e_x\nabla_y)\delta\eta\nonumber\\
&
+\eta_{\rm e}\nabla^2 \bm V + (\zeta_{\rm e}+(1-2/d)\eta_{\rm e})\nabla Z,\label{eq:lv}
\end{align}
where $Z=\nabla\cdot\bm V$, $\delta\eta=\eta_\psi\psi$ is the change in the viscosity with $\eta_\psi=(\p\eta/\p\phi)|_{\phi=\phi_{\rm e}}$, and $\kappa'=(\p^2 f/ \p \phi^2)|_{\phi=\phi_{\rm e}}=\kappa+2b_3\phi_{\rm e}+3b_4\phi_{\rm e}^2$. The shear and bulk viscosities of the homogeneous state are $\eta_{\rm e}=\eta(\phi_{\rm e})$ and $\zeta_{\rm e}=\zeta(\phi_{\rm e})$.
Taking divergence in the last equation yields
\begin{align}
\rho_{\rm e}\bigg(\pp{}{t}+\dot{\gamma}y\pp{}{x}\bigg)Z
&=
-2\rho_{\rm e}\dot{\gamma}\nabla_x V_y+(1+m_{\rm e})\nabla^2[\alpha\psi-\delta m/K_0]\nonumber\\
&+2\dot{\gamma}\nabla_x\nabla_y\delta\eta+\eta_{\rm e}'\nabla^2 Z,
\label{eq:Z}
\end{align}
where $\eta_{\rm e}'=\zeta_{\rm e}+(2-2/d)\eta_{\rm e}$. A closed set of equations for $\delta m$, $\psi$, and $Z$ is obtained by neglecting the first term on the r.h.s. of Eq.~(\ref{eq:Z}), which couples the shear flow to the transverse velocity (see Appendix~\ref{app:transverse} for the role of this term). In Fourier space, the equations read
\begin{widetext}
\be
(\pp{}{t}-\dot{\gamma}q_x\pp{}{q_y})
\left(
\begin{array}{c}
m_{\bm q} \\
\psi_{\bm q} \\
Z_{\bm q}
\end{array}
\right)
=
\left(
\begin{array}{ccc}
0 & 0 & -(1+m_{\rm e}) \\
L\alpha & -L(\kappa'+Dq^2) & 0 \\
q^2/K_0'\rho_{\rm e} & -(\alpha' q^2+2\dot{\gamma}q_xq_y\eta_\psi)/\rho_{\rm e} & -q^2\eta_{\rm e}'/\rho_{\rm e}
\end{array}
\right)
\left(
\begin{array}{c}
m_{\bm q} \\
\psi_{\bm q} \\
Z_{\bm q}
\end{array}
\right),
\label{eq:matrix}
\en
\end{widetext}
where $\alpha'=(1+m_{\rm e})\alpha$ and $K_0'=K_0/(1+m_{\rm e})$. The stability is first examined for a fixed wavevector, i.e., without the advection term on the l.h.s. of Eq.~(\ref{eq:matrix}), and in the long-wavelength limit. The shear-independent and shear-dependent parts of the matrix elements are denoted by $A_{ij}$ and $B_{ij}$, respectively:
$A_{13}=1+m_{\rm e}$,
$A_{21}=L\alpha$,
$A_{22}=L\kappa'$,
$A_{31}=q^2/K_0'\rho_{\rm e}$,
$A_{32}=\alpha' q^2/\rho_{\rm e}$,
$A_{33}=q^2\eta_{\rm e}'/\rho_{\rm e}$,
and $B_{32}=2\dot{\gamma}q_xq_y\eta_\psi/\rho_{\rm e}$.
The element $B_{32}$ changes its sign with the direction of the wavevector.
The characteristic equation reads
\be
\lambda^3+a_2\lambda^2+a_1\lambda+a_0=0,\label{eq:chara}
\en
where $a_2=A_{22}+A_{33}$, $a_1=A_{22}A_{33}+A_{13}A_{31}$, and $a_0=A_{13}[A_{22}A_{31}-A_{21}(A_{32}+B_{32})]$.
Because $a_2>0$ and $a_1>0$, this equation has a positive root when $a_0<0$, i.e., when $\dot{\gamma}\alpha\eta_\psi\sin2\theta>(1+m_{\rm e})[\kappa'/K_0-\alpha^2]$, where $\theta$ is the angle of $\bm q$ relative to the $q_x$ axis. In the directions $\sin2\theta=\pm1$, where the left-hand side is largest, this condition reads
\be
\dot{\gamma}|\alpha\eta_\psi| > (1+m_{\rm e})\bigg[\frac{\kappa'}{K_0}-\alpha^2\bigg] >0,\label{eq:instability}
\en
where the second inequality ensures thermodynamic stability: the spinodal point is $\kappa'=K_0\alpha^2$ (see Appendix~\ref{app:isobaric} for the NPT case).
Modes with a fixed wavevector thus become unstable for $\dot{\gamma}>\dot{\gamma}_{\rm c}$, where
\be
\dot{\gamma}_{\rm c}=\frac{1+m_{\rm e}}{|\alpha\eta_\psi|}\bigg[\frac{\kappa'}{K_0}-\alpha^2\bigg].\label{eq:gammac}
\en
The instability first appears for $q_x=q_y$ when $\alpha\eta_\psi>0$ and for $q_x=-q_y$ when $\alpha\eta_\psi<0$. For example, TPP is the former case ($\alpha>0$ and $\eta_\psi>0$), and water is the latter ($\alpha<0$ and $\eta_\psi>0$).

The mechanism of the instability is as follows. A fluctuation of $\psi$ changes the viscosity by $\eta_\psi\psi$. Under shear, the viscous stress $\eta\dot{\gamma}$ becomes inhomogeneous, and its divergence drives a compressional flow for the wavevectors with $q_xq_y\neq0$ due to the term $2\dot{\gamma}\nabla_x\nabla_y\delta\eta$. Then the density change induced by the compressional flow changes $\psi$ through the coupling $-\alpha\phi m$ in Eq.~(\ref{eq:f}). For $\alpha\eta_\psi q_xq_y>0$, the density change increases $\psi$, and the feedback is positive. The feedback is proportional to $\dot{\gamma}\alpha\eta_\psi$ and competes with thermal relaxation, which is proportional to the inverse susceptibility $\kappa'-K_0\alpha^2$. The critical shear rate $\dot{\gamma}_{\rm c}$ is the shear rate at which the two balance in the most favorable directions, $|q_x|=|q_y|$. 
%The critical shear rate therefore vanishes at the spinodal point.
%The instability is thus driven by the product of the density contrast $\alpha$ and the viscosity contrast $\eta_\psi$: the order parameter fluctuation modulates the viscosity, the resulting inhomogeneous shear stress drives a compressive flow, and the density change feeds back on $\phi$ through $\alpha$.

\subsection{Wavevector advection}
\label{sec:advection}
The condition $\dot{\gamma}>\dot{\gamma}_{\rm c}$ of Sec.~\ref{sec:instability} was derived for a fixed wavevector $\bm q$. In shear flow, however, the wavevector of each Fourier mode is advected as~\cite{onuki1979nonequilibrium,Onukibook}
\be
\bm q(t)=\big(q_x, q_y(0)-\dot{\gamma}q_x t\big),
\label{eq:appA-lagq}
\en
so that a mode initially in the unstable directions migrates outside them after a finite time and eventually decays. Since the growth rate of the unstable root of Eq.~(\ref{eq:chara}), $\lambda\simeq -a_0/a_1$, is proportional to the kinetic coefficient $L$ for small $L$, the amplification accumulated during this time remains finite and is small for small $L$. The instability at a fixed wavevector therefore does not indicate a large growth, and the total amplification of a fluctuation along its trajectory needs to be calculated.

The amplification is calculated by eliminating the fast variables. Along the trajectory (\ref{eq:appA-lagq}) the linearized equation becomes $\dot{\bm X}=M(\bm q(t))\bm X$, where $\bm X=(m_{\bm q},\psi_{\bm q},Z_{\bm q})$ and the dot represents the advective derivative on the l.h.s. of Eq.~(\ref{eq:matrix}).  The order parameter $\psi$ evolves at rates ${\cal O}(L)$, whereas the pair $(m,Z)$ relaxes through the acoustic and viscous rates $c_{\rm s}q$ and $q^2\eta_{\rm e}'/\rho_{\rm e}$, where $c_{\rm s}$ is the sound speed. To the lowest order in $L$, the elimination of the fast variables $m$ and $Z$ yields $Z=0$ and
\be
\frac{m}{\psi}=K_0'(\alpha'+\dot{\gamma}w\eta_\psi),
\label{eq:appA-chi3}
\en
where $w=2q_xq_y/q^2=\sin2\theta$. The amplitude of $\psi$ then follows $\dot\psi=\lambda_\psi\psi$ with
\be
\lambda_\psi=L\big[\alpha K_0'(\alpha'+\dot{\gamma}w\eta_\psi)-\kappa'-Dq^2\big].
\label{eq:appA-lameff0}
\en
Using $\dot{\gamma}_{\rm c}$, Eq.~(\ref{eq:appA-lameff0}) in the long-wavelength limit reads
\be
\lambda_\psi=\Lambda\dot{\gamma}(w-r),
\label{eq:appA-lameff}
\en
for $\alpha\eta_\psi>0$, where $r\equiv \dot{\gamma}_{\rm c} / \dot{\gamma}$. For $\alpha\eta_\psi<0$, $w$ is replaced by $-w$. The dimensionless number $\Lambda$ is given by
\be
\Lambda=L K_0'|\alpha\eta_\psi|=\frac{|\lambda_0|}{\dot{\gamma}_{\rm c}},
\label{eq:appA-Lambda}
\en
where $\lambda_0=-L(\kappa'-K_0\alpha^2)$ is the zero-shear relaxation rate of the order parameter fluctuation in the long-wavelength limit. For finite $L$, $\Lambda$ is replaced with $\Lambda=|\lambda_0|/[(1+\delta_L)\dot{\gamma}_{\rm c}]$ with $\delta_L=L\kappa'/\Gamma_\rho$ being the leading correction, where $\Gamma_\rho=(1+m_{\rm e})/K_0'\eta_{\rm e}'$ is the rate at which the density relaxes against the viscous stress (see Appendix~\ref{app:deltaL}). Thus the instability is governed by the two dimensionless parameters: inverse shear rate $r$ and thermodynamic relaxation rate $\Lambda$.
%have the following physical meaning. Equation~(\ref{eq:appA-lameff}) can be rewritten as
%\be
%\lambda_\psi=-\frac{L(\kappa'-K_0\alpha^2)}{1+\delta_L}\Big(1-\frac{\dot{\gamma}}{\dot{\gamma}_{\rm c}}\sin2\theta\Big).
%\en
%Shear thus acts on a fluctuation with the wavevector direction $\theta$ as a change of the curvature of the free energy from $\kappa'-K_0\alpha^2$ to $(\kappa'-K_0\alpha^2)[1-(\dot{\gamma}/\dot{\gamma}_{\rm c})\sin2\theta]$, which is called the effective curvature below. For $\dot{\gamma}>\dot{\gamma}_{\rm c}$, the effective curvature is negative in a range of directions, as the curvature is in the spinodal region. The ratio $\dot{\gamma}/\dot{\gamma}_{\rm c}$ thus measures the change of the curvature by shear. The number $\Lambda$ concerns the rotation of the wavevector. The shear-dependent part of $\lambda_\psi$ is $\Lambda\dot{\gamma}w$, and the direction of the wavevector changes at the rate $\dot{\gamma}$ [Eq.~(\ref{eq:appA-lagq})]. The number $\Lambda$ is the ratio of the two rates. For $\Lambda\ll1$, the direction changes before the amplitude of the fluctuation responds, and the effect of shear on the amplitude is small. For $\Lambda\gtrsim1$, the amplitude follows the change of the direction. The effect of shear on a fluctuation is therefore determined by $\dot{\gamma}/\dot{\gamma}_{\rm c}$ and $\Lambda$.

\begin{figure}[t]
\centering
\includegraphics[width=85mm]{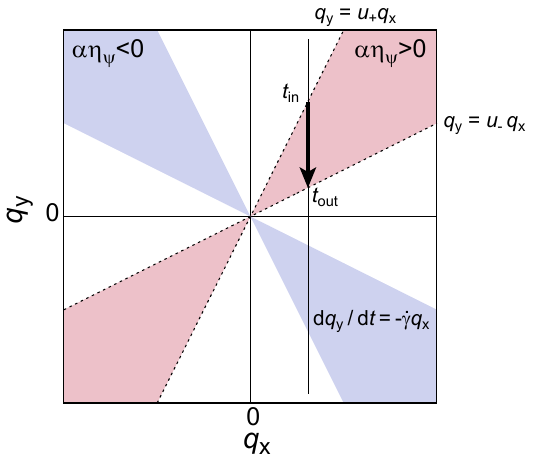}
\caption{
Unstable region of the order-parameter fluctuation in the wavevector plane, $\lambda_\psi>0$ in the long-wavelength limit, drawn for $r=\dot{\gamma}_{\rm c}/\dot{\gamma}=0.8$. The red regions are for $\alpha\eta_\psi>0$ and the blue regions are for $\alpha\eta_\psi<0$. The dotted lines are the boundaries $q_y=u_\pm q_x$ for $\alpha\eta_\psi>0$. The vertical line is the trajectory of a wavevector advected by the shear flow, Eq.~(\ref{eq:appA-lagq}). The mode grows from $t_{\rm in}$ to $t_{\rm out}$, as indicated by the arrow.
}
\label{fig::unstable}
\end{figure}

The directions of the wavevector in which a fluctuation grows are obtained from Eq.~(\ref{eq:appA-lameff}). The growth condition $\lambda_\psi>0$ is $w>r$ and $r<1$. With $u=q_y/q_x=\tan\theta$, this condition is satisfied for $u_-<u<u_+$, where $u_\pm=(1\pm\sqrt{1-r^2}) / r$. The unstable region in the wavevector plane is shown in Fig.~\ref{fig::unstable}. For $\alpha\eta_\psi>0$, the region corresponds to the angular range $\theta_-<\theta<\theta_+$, with $\theta_-=\frac{1}{2}\sin^{-1}r$ and $\theta_+=\pi/2-\theta_-$, and the opposite range obtained by $\bm q\to-\bm q$, as shown in the red region in Fig.~\ref{fig::unstable}. The unstable region appears at $\theta=\pi/4$ for $\dot{\gamma}=\dot{\gamma}_{\rm c}$ and becomes broader as $\dot{\gamma}$ increases. For $\alpha\eta_\psi<0$, the unstable region appears in the blue region in Fig.~\ref{fig::unstable}.

A fluctuation grows only while its wavevector is in the unstable region. Along the trajectory (\ref{eq:appA-lagq}), $du/dt=-\dot{\gamma}$, so that a mode moves along a vertical line in Fig.~\ref{fig::unstable} and is unstable from $t=t_{\rm in}$, at which $u=u_+$, to $t=t_{\rm out}$, at which $u=u_-$. The amplification factor $G$ of a fluctuation during this passage is defined as
\be
\ln G=\int_{t_{\rm in}}^{t_{\rm out}} dt \lambda_\psi=\int_{u_-}^{u_+} du\frac{\lambda_\psi}{\dot{\gamma}}=\Lambda\Phi(r),
\label{eq:lnG}
\en
where
\be
\Phi(r)=-r(u_+-u_-)+\ln\frac{u_+^2+1}{u_-^2+1}.
\label{eq:appA-Phib}
\en
Because $G$ is the largest amplification along the trajectory, for small $G$ the transition does not proceed by the deterministic growth of a single mode even when $\dot{\gamma}>\dot{\gamma}_{\rm c}$.

\subsection{Structure factor in the steady state}
\label{sec:sk-theory}
Even when $G$ is small and the phase transition is not induced deterministically, shear changes the thermal fluctuations of the metastable liquid, which is manifested in the structure factor of the order parameter, $S_\phi(\bm q)=\av{|\psi_{\bm q}|^2}/V$, where $V$ is the system volume. To calculate $S_\phi$, a random force in Eq.~(\ref{eq:fluc1}) is incorporated as $\dot{\psi}_{\bm q}=\lambda_\psi\psi_{\bm q}+\theta_{\bm q}'$, where $\lambda_\psi$ includes the finite-$L$ correction, Eq.~(\ref{eq:appC-lam}), and $\theta_{\bm q}'$ is a Gaussian white noise. Its variance is $2LV_0^2/(1+\delta_L)$, which is fixed by the requirement that the steady state without shear reproduces the equilibrium structure factor. The structure factor then obeys
\be
\bigg(\pp{}{t}-\dot{\gamma}q_x\pp{}{q_y}\bigg)S_\phi(\bm q)=2\lambda_\psi(\bm q)S_\phi(\bm q)+\frac{2LV_0^2}{1+\delta_L},
\label{eq:Sk-kinetic}
\en
where $V_0^2=v_0/\ell_0^d$ is the noise amplitude and $\delta_L$ is the finite-$L$ correction (see Appendix~\ref{app:deltaL}). Without shear, the steady solution has the Ornstein--Zernike form
\be
S_\phi^{(0)}(q)=\frac{V_0^2}{\kappa'-K_0\alpha^2+Dq^2}.
\label{eq:Sk-OZ}
\en
Under shear, the steady solution for $q_x>0$ is obtained by integrating Eq.~(\ref{eq:Sk-kinetic}) along the trajectory (\ref{eq:appA-lagq}) as
\be
S_\phi(\bm q)=\frac{2LV_0^2}{(1+\delta_L)\dot{\gamma}}\int_0^\infty ds
\exp\bigg[\frac{2}{\dot{\gamma}}\int_u^{u+s}du'\lambda_\psi(q_x,u'q_x)\bigg],
\label{eq:Sk-steady}
\en
where $\lambda_\psi(q_x,q_y)$ is evaluated on the trajectory, and $u+s$ is the value of $q_y/q_x$ at the time $s/\dot{\gamma}$ before the wavevector reaches $\bm q$. In the long-wavelength limit, Eq.~(\ref{eq:appA-lameff0}) yields
\be
\frac{S_\phi(\bm q)}{S_\phi^{(0)}(0)}=2\Lambda r\int_0^\infty ds
\bigg[\frac{1+(u+s)^2}{1+u^2}\bigg]^{\pm2\Lambda}e^{-2\Lambda rs},
\label{eq:Sk-long}
\en
where the plus (minus) sign is for $\alpha\eta_\psi>0$ ($<0$). 
%The right-hand side depends only on the direction of the wavevector, $u=\tan\theta$, and on $\Lambda$ and $r$.

Equations~(\ref{eq:Sk-steady})--(\ref{eq:Sk-long}) have three consequences, which are compared with the numerical simulations in Sec.~\ref{sec:results}. 
First, the instantaneous shear contribution to $\lambda_\psi$ increases or decreases the growth rate depending on the sign of $\alpha\eta_\psi q_xq_y$. This coupling results in anisotropic $S_\phi$ even for $\dot{\gamma}<\dot{\gamma}_{\rm c}$.
%First, at small $q$, $S_\phi$ is larger than $S_\phi^{(0)}$ in the directions where shear slows the relaxation, i.e., where $\lambda_\psi$ exceeds its value without shear. These are the directions $q_xq_y>0$ for $\alpha\eta_\psi>0$ and $q_xq_y<0$ for $\alpha\eta_\psi<0$. In the opposite directions, $S_\phi$ is smaller than $S_\phi^{(0)}$. Because $\lambda_\psi$ depends on the direction of $\bm q$ for any $\dot{\gamma}$, $S_\phi$ is anisotropic also for $\dot{\gamma}<\dot{\gamma}_{\rm c}$. 
Second, $S_\phi(\bm q)$ is the result of the growth and decay accumulated along the trajectory, on which the wavevector changes as $d\theta/dt=-\dot{\gamma}\cos^2\theta$. The directions in which $S_\phi$ is maximal and minimal are therefore displaced clockwise from $\theta=\pm\pi/4$, at which $|w|$ is maximal. 
%Third, the last term of Eq.~(\ref{eq:Sk-exponent}) weakens the effect of shear with increasing $q$, and $S_\phi$ approaches $S_\phi^{(0)}$ at large $q$. The anisotropy is thus largest at small $q$, with the orientation determined by the sign of $\alpha\eta_\psi$ and displaced by the rotation of the wavevector.
Third, the patterns of $S_\phi$ are not symmetric with respect to the sign of $\alpha\eta_\psi$, because the wavevector rotates clockwise for both signs. For $\alpha\eta_\psi>0$ ($<0$), $\lambda_\psi$ is largest at $u=1$ ($u=-1$), and the wavevector then rotates toward the $q_x$ ($q_y$) axis. Because $d\theta/du=\cos^2\theta$, the $\theta$ range where the enhancement occurs becomes broad near the $q_x$ axis and narrow near the $q_y$ axis. The enhancement is therefore spread over a wide range of angles for $\alpha\eta_\psi>0$ and is confined to a narrow range near the $q_y$ axis for $\alpha\eta_\psi<0$. 
%For $\Lambda=0.4$ and $r=1$, the maximum of $S_\phi/S_\phi^{(0)}(0)$ is 3.1 at $\theta=21^\circ$ for $\alpha\eta_\psi>0$ and 2.2 at $\theta=107^\circ$ for $\alpha\eta_\psi<0$, and the angular range in which $S_\phi/S_\phi^{(0)}(0)-1$ exceeds half of its maximum is $87^\circ$ and $26^\circ$ wide, respectively. For $\alpha\eta_\psi<0$, in addition, $q=|q_x|(1+u^2)^{1/2}$ increases on the downstream side, so that the fluctuations amplified at small $q$ are advected to large $q$, and the enhancement extends to large $q$.

%Fifth, the magnitude of the enhancement depends on $\Lambda$ differently below and above $\dot{\gamma}_{\rm c}$. For $\dot{\gamma}<\dot{\gamma}_{\rm c}$, $\lambda_\psi$ is negative in all the directions, and $S_\phi\le S_\phi^{(0)}(0)/(1-\dot{\gamma}/\dot{\gamma}_{\rm c})$ for any $\Lambda$. For $\dot{\gamma}>\dot{\gamma}_{\rm c}$, the fluctuations grow in the unstable region, and the enhancement increases steeply with $\Lambda$; it remains finite because the rotation of the wavevector terminates the growth. As an example, the maximum of $S_\phi/S_\phi^{(0)}(0)$ over the directions is 2.1, 2.6, 3.5, and 4.2 for $\Lambda=0.2$, $0.4$, $1$, and $2$ at $\dot{\gamma}=0.83\dot{\gamma}_{\rm c}$, where the bound is 6, and is 3.6, 7.1, 33, and 280 at $\dot{\gamma}=2\dot{\gamma}_{\rm c}$. In the long-wavelength limit, the enhancement is thus large when $\dot{\gamma}>\dot{\gamma}_{\rm c}$ and $\Lambda$ is of order unity or larger.

\subsection{Limiting behavior}
\label{app:transient-limits}
The dependence of $G$ on the shear rate in two limiting cases reads
\begin{align}
\dot{\gamma}\searrow\dot{\gamma}_{\rm c}:\quad&
\Phi(r)\simeq\tfrac{2}{3}(1-r^2)^{3/2}
\;\;\Rightarrow\;\;
\ln G\simeq\frac{2\Lambda}{3}
\Big[\frac{2(\dot{\gamma}-\dot{\gamma}_{\rm c})}{\dot{\gamma}}\Big]^{3/2},
\label{eq:appA-onset}\\
\dot{\gamma}\gg\dot{\gamma}_{\rm c}:\quad&
\Phi(r)\simeq2\ln\frac{2\dot{\gamma}}{\dot{\gamma}_{\rm c}}-2
\;\;\Rightarrow\;\;
G\simeq
e^{-2\Lambda}\Big(\frac{2\dot{\gamma}}{\dot{\gamma}_{\rm c}}\Big)^{2\Lambda}.
\label{eq:appA-strong}
\end{align}
Equation~(\ref{eq:appA-onset}) represents a power-law onset with exponent $3/2$. At high shear rates, the growth rate increases in proportion to $\dot{\gamma}$ from Eq.~(\ref{eq:appA-lameff}), whereas the time during which the direction of the wavevector is favorable decreases as $1/\dot{\gamma}$. The two dependences cancel, and $\ln G$ increases only logarithmically with $\dot{\gamma}$. The amplification factor $G$ is therefore of order unity unless $\Lambda$ is large. For $\Lambda=0.2$, which corresponds to $L=0.01$, $G=1.2$, $1.7$, $2.3$, and $3.0$ at $\dot{\gamma}/\dot{\gamma}_{\rm c}=2$, $5$, $10$, and $20$. %This implies that $L$ needs to be larger to induce phase transition under shear.

The linear theory thus gives the following picture of the effect of shear on the metastable liquid. Thermal fluctuations are continuously generated by the noise, and their amplitudes change while their wavevectors are rotated by the flow. 
%In the steady state the fluctuations are enhanced anisotropically, and the enhancement becomes larger as increasing $\dot{\gamma}$ and $\Lambda$. Because the nucleation rate depends exponentially on the amplitude of the fluctuations~\cite{langer1992introduction,Debenedetti}, the enhancement is expected to promote the nucleation of liquid~2. The relaxation rate of the order parameter $L$ as well as the shear rate and the noise amplitude is essential for inducing LLT under shear flow.
In the steady state the fluctuations are enhanced anisotropically, and the enhancement becomes larger with increasing $\dot{\gamma}$ and $\Lambda$. From Eq.~(\ref{eq:appA-lameff}), shear reduces $\kappa'-K_0\alpha^2$ to $(\kappa'-K_0\alpha^2)(1-\dot{\gamma}w/\dot{\gamma}_{\rm c})$, and the barrier for nucleation decreases. Because the nucleation rate depends exponentially on the barrier~\cite{langer1992introduction,Debenedetti}, the enhancement is expected to promote the nucleation of liquid~2. Inducing the LLT under shear thus requires, in addition to $\dot{\gamma}\gtrsim\dot{\gamma}_{\rm c}$, a kinetic coefficient $L$ large enough for $\Lambda\gtrsim1$ and the thermal noise that initiates the nucleation.

\section{Numerical results}
\label{sec:numerical}
\subsection{Methods}
\label{sec:methods}
\subsubsection{Imposing shear flow}
The nonlinear hydrodynamic equations Eqs.~(\ref{eq:cont})--(\ref{eq:TDGL}) with the noise terms (\ref{eq:fluc1})--(\ref{eq:fluc2}) are integrated on a fixed square lattice in the oblique coordinates $\hat x=x-\gamma(t)(y-y_0)$, $\hat y=y$, with $\gamma(t)=\int_0^t dt'\dot{\gamma}(t')$, following the deformed-coordinate method~\cite{rogallo1981numerical,onuki1997new,kobayashi2011implementation}. Periodicity in the oblique frame is equivalent to Lees--Edwards boundary conditions in the laboratory frame. 
The mean shear flow $\dot{\gamma}(y-y_0)\bm e_x$ is absorbed in the coordinate transformation, and $\bm v$ in this subsection denotes the deviation of the velocity from it. Velocity and momentum are stored as contravariant components of this deviation: $\hat v^1=v_x-\gamma v_y$ and $\hat v^2=v_y$. The momentum equation for $\hat g^1=\rho\hat v^1$ then has the term $-2\dot{\gamma}\hat g^2$, where one factor of $\dot{\gamma}$ comes from the advection by the mean flow, $v_y\partial_y(\dot{\gamma}y)$, and the other from the time dependence of the basis, $d\gamma/dt$ in $\hat v^1$.
%Velocity and momentum are stored as contravariant components: $\hat v^1=v_x-\gamma v_y$ and $\hat v^2=v_y$. The momentum equation for $\hat g^1=\rho\hat v^1$ then has the term $-2\dot{\gamma}\hat g^2$.
%, one factor $\dot{\gamma}$ from the mean-shear stretching of the momentum and one from the rotation of the basis vectors. 
Stresses are evaluated in contravariant form, $\hat\sigma^{ij}=\eta(G^{in}\p_n\hat v^j+G^{jm}\p_m\hat v^i)$ etc., with the inverse metric $G^{11}=1+\gamma^2$, $G^{12}=-\gamma$, $G^{22}=1$. The lattice is remapped to zero strain each time $\gamma$ reaches unity. The random stresses are generated as laboratory-frame white noises satisfying Eq.~(\ref{eq:fluc2}) and transformed to contravariant components, which reproduces the required covariances including the $\gamma$-dependent cross correlations.

\subsubsection{Isobaric simulations}
Because the two liquids differ in density, the numerical simulations are performed at constant pressure rather than constant volume~\cite{takae2020role}. The system volume $V$ is treated as a dynamical variable relaxing toward the minimum of the Gibbs potential $\mathcal{G}=\mathcal{F}+P_0V$,
\be
\dot V=-L_V\pp{\mathcal{G}}{V}=L_V(P-P_0),
\label{eq:isobaric}
\en
where $P$ is the spatial average of the local pressure Eq.~(\ref{eq:pre}) and $P_0$ is the external pressure. The volume change is realized by an isotropic affine rescaling of the grid, $\Delta x\to(1+\ve\Delta t/d)\Delta x$ with $\ve=\dot V/V$, and the velocity is divided into the affine part $\bm v_{\rm a}=\ve\bm r/d$ and a non-affine contribution $\bm v_{\rm na}$. Written in terms of the semi-Lagrangian derivative $D_{\rm a}/D_{\rm a}t=\p_t+\bm v_{\rm a}\cdot\nabla$, the equations of motion for $\rho$, $\rho\bm v_{\rm na}$, and $\phi$ read~\cite{takae2020role}
\begin{align}
\frac{D_{\rm a}}{D_{\rm a}t}\rho &=-\nabla\cdot \rho\bm v_{\rm na}-\ve \rho,\\
\frac{D_{\rm a}}{D_{\rm a}t}\rho\bm v_{\rm na}&=-\nabla\cdot \rho\bm v_{\rm na}\bm v_{\rm na}-\nabla\cdot\tensor{\Pi}+\nabla\cdot(\tensor{\sigma}+\tensor{\sigma}^{\rm R}) -\ve \rho\bm v_{\rm na},\\
\frac{D_{\rm a}}{D_{\rm a}t}\phi &=-\bm v_{\rm na}\cdot \nabla\phi-L\Gamma + \theta.
\end{align}
Since the rescaling is isotropic, it commutes with the oblique-coordinate shear transformation, so that the accumulated strain $\gamma(t)$ remains unchanged. The values $P_0=0$ and $L_V=V$ are used.
%The isobaric reference state and spinodal that result are given in Appendix~\ref{app:isobaric}.

\subsubsection{Numerical scheme and parameters}
\label{sec:numparams}
The equations are discretized on a collocated square lattice of $256\times256$ sites with $\Delta x=\ell_0$ in the oblique coordinates, with periodic boundary conditions. The time integration is the explicit Euler scheme with $\Delta t=10^{-3}$ for $\dot{\gamma}\le0.1$ and $\Delta t=10^{-4}/\dot{\gamma}$ for larger $\dot{\gamma}$. Each run starts from the homogeneous liquid~1 at $\tau=0.1$, where liquid~1 is thermodynamically stable. The system is equilibrated without shear, and is then quenched to the target $\tau$ and sheared at the same time, which is set to $t=0$.

The parameters of the free energy are those of Sec.~\ref{sec:GL} with $K_0=0.04$ and $D=2$ for $\alpha=\pm 10$. The viscosities are $\eta(\phi)=1+50(\phi+0.3)$ and $\zeta=(2/d)\eta$~\cite{takae2020role}, so that $\eta_\psi=50$ and $\eta_{\rm e}=\eta(\phi_{\rm e})\simeq9.5$ for $\phi_{\rm e}\simeq-0.13$. The equations are integrated in the dimensionless form of Ref.~\cite{takae2020role}, with the units being $\ell_0$ (length), $t_0$ (time), $m_0$ (mass), $k_{\rm B}T_0$ (energy), and $v_0$ (volume). The unit length $\ell_0$ in Eq.~(\ref{eq:F}) is set equal to the lattice spacing $\Delta x$. The unit time is $t_0=\ell_0^2/\nu_0$, where $\nu_0$ is the kinematic viscosity of liquid~1 at the reference temperature. The density is measured in units of $\rho_0=m_0/v_0$ with $m_0$ being the molecular mass. The dimensionless mass $M=m_0\ell_0^2/(t_0^2k_{\rm B}T_0)$ and the noise amplitude $V_0=(v_0/\ell_0^d)^{1/2}$ in Eqs.~(\ref{eq:mom}) and (\ref{eq:TDGL}) are set to $M=1$ and $V_0=0.01$ for numerical convenience, while the values estimated for TPP are $M=2.8\times10^{11}$ and $V_0=2\times10^{-7}$. Neither $M$ nor $V_0$ enters $\dot{\gamma}_{\rm c}$, $\Lambda$, and $G$ of the linear theory: $M$ affects only the fast relaxation of the density and the velocity, and $V_0$ sets the rate of the thermal nucleation. To examine the role of the noise amplitude, $V_0=0.02$ is also adopted in Sec.~\ref{sec:results}.

%We assume $\zeta=2\eta/d$ for simplicity. The kinetic coefficient $L$ and the shear rate $\dot{\gamma}$ are the control parameters. The amplitude of the thermal noise terms $\theta$ and $\sigma^{\rm R}$ is given by the dimensionless molecular volume~\cite{takae2020role}: $V_0=(v_0/\ell_0^d)^{1/2}$. Though the value estimated for TPP is $2\times10^{-7}$, we adopt $V_0=0.01$ and $0.02$ for numerical convenience.

The isobaric reference states used for the linear theory (Appendix~\ref{app:isobaric}) are characterized by
\begin{center}
\begin{tabular}{c|cccc}
\multicolumn{5}{c}{$\alpha=10$}\\
$\tau$ & $\kappa'-K_0\alpha^2$ & $\dot{\gamma}_{\rm c}$ & $K_0'|\alpha\eta_\psi|$ & $\delta_L/L$\\
\hline
$-0.123$ & $0.630$ & $0.0298$ & $21.1$ & $3.85$\\
$-0.124$ & $0.415$ & $0.0197$ & $21.1$ & $3.73$\\
\end{tabular}
\\[2ex]
\begin{tabular}{c|cccc}
\multicolumn{5}{c}{$\alpha=-10$}\\
$\tau$ & $\kappa'-K_0\alpha^2$ & $\dot{\gamma}_{\rm c}$ & $K_0'|\alpha\eta_\psi|$ & $\delta_L/L$\\
\hline
$-0.129$ & $0.638$ & $0.0336$ & $19.0$ & $3.10$\\
$-0.130$ & $0.457$ & $0.0241$ & $19.0$ & $3.03$\\
\end{tabular}
\end{center}
where $\dot{\gamma}_{\rm c}$ is evaluated from Eq.~(\ref{eq:gammac}). For $L=0.02$, $0.05$, $0.1$, and $0.2$, the dimensionless parameter $\Lambda$ takes the value $\Lambda=0.39$, $0.89$, $1.5$, and $2.4$ for $\alpha=10$, and $\Lambda=0.36$, $0.82$, $1.5$, and $2.4$ for $\alpha=-10$. The temperatures of the shear simulations, $\tau=-0.123$ and $-0.124$ ($\alpha=10$) and $\tau=-0.129$ and $-0.130$ ($\alpha=-10$), are thus metastable but close to the respective spinodal points, $\tau_{\rm s}=-0.1248$ and $-0.1311$ (see Appendix~\ref{app:isobaric}). The quench depth is incorporated in $|\lambda_0|$ and $\dot{\gamma}_{\rm c}$ in the same way and
cancels from $\Lambda$ apart from the weak dependence of $\delta_L$ on $\kappa'$ (see Appendix~\ref{app:deltaL}). For $\tau\le-0.125$ ($\alpha=10$) and $\tau\le-0.132$ ($\alpha=-10$) no metastable liquid-1 branch exists at $P_0=0$, while it is not easy to distinguish nucleation growth and spinodal growth by the observed patterns due to the crossover nature of the first-order phase transition under thermal fluctuation~\cite{binder1987theory}.

\subsection{Results}
\label{sec:results}

For each set of $\alpha$, $\tau$, $L$, $\dot{\gamma}$, and $V_0$, six independent runs are performed. The fraction of liquid~2 is defined as the fraction of lattice sites with $\phi>0.2$. The incubation time $t_{\rm inc}$ is defined as the time at which this fraction first exceeds $0.001$. When the fraction remains below this value until $t=10^4$, the run is counted as a run without transition.

\begin{figure}[t]
\centering
\includegraphics[width=85mm]{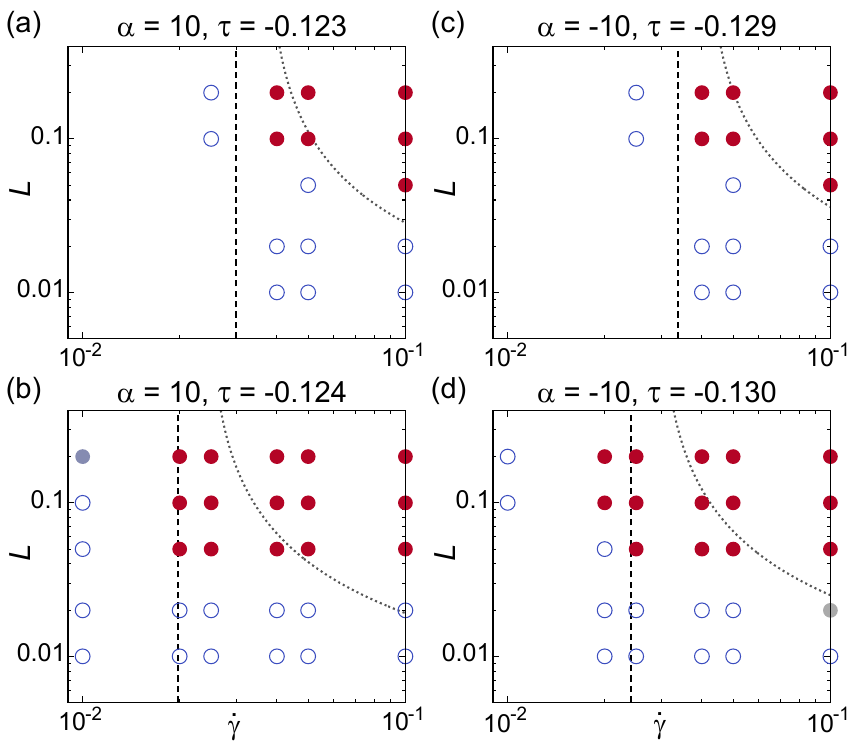}
\caption{
Occurrence of the transition in the $(\dot{\gamma},L)$ plane for $V_0=0.01$. (a) $\alpha=10$, $\tau=-0.123$. (b) $\alpha=10$, $\tau=-0.124$. (c) $\alpha=-10$, $\tau=-0.129$. (d) $\alpha=-10$, $\tau=-0.130$. Filled circles: the transition occurs in all the runs. Open circles: the transition does not occur. Shaded circles: the transition occurs in part of the runs. The vertical dashed lines indicate $\dot{\gamma}_{\rm c}$ in Eq.~(\ref{eq:gammac}), and the dotted curves represent $\ln G=\Lambda\Phi(r)=1$ in the linear theory.
}
\label{fig::diagram}
\end{figure}

\subsubsection{Shear-induced phase transition}
The first question is under which conditions shear induces the transition in metastable liquid~1. The occurrence of the transition is plotted in the $(\dot{\gamma},L)$ plane for $V_0=0.01$ in Fig.~\ref{fig::diagram}. For $L\ge0.1$, the lowest shear rate at which the transition occurs agrees with $\dot{\gamma}_{\rm c}$ of Eq.~(\ref{eq:gammac}). In Fig.~\ref{fig::diagram}(d), the transition occurs also at $\dot{\gamma}=0.02\simeq 0.83\dot{\gamma}_{\rm c}$. At $\tau=-0.124$, $L=0.2$, and $\dot{\gamma}=0.01\simeq 0.5\dot{\gamma}_{\rm c}$, only one run exhibits the transition: the transition is a rare nucleation event caused by the thermal fluctuations. For $L\ge0.1$, $\dot{\gamma}_{\rm c}$ of the linear theory thus gives the shear rate at which the transition is induced. The transition at $0.83\dot{\gamma}_{\rm c}$ in Fig.~\ref{fig::diagram}(d) is of a different origin from the rare nucleation at $0.5\dot{\gamma}_{\rm c}$: the nonlinear fluctuations shift the average order parameter and lower $\dot{\gamma}_{\rm c}$ itself, as shown in Sec.~\ref{sec:sk-results}.

However, a shear rate above $\dot{\gamma}_{\rm c}$ is not a sufficient condition for the transition. The transition occurs only when $L$ is large. For $L\le 0.02$, liquid~1 remains metastable up to the highest shear rate. For small $L$, the dimensionless parameter $\Lambda$ and accordingly the amplification factor $G$ remain small: $\ln G \lesssim 1$ even for the highest shear rate for $L\le 0.02$, as shown in Fig.~\ref{fig::diagram}.

The requirement of both large $\dot{\gamma}$ and large $\Lambda$ is consistent with the linear theory of Sec.~\ref{sec:theory}. The transition observed for $L\ge0.05$ at $\dot{\gamma}\gtrsim\dot{\gamma}_{\rm c}$, where $\ln G<1$, is due to the thermal noise imposed in the simulations: the barrier for nucleation, which decreases as the spinodal point is approached, also sets the threshold. The shear-induced transition thus requires $\dot{\gamma}\gtrsim\dot{\gamma}_{\rm c}$ and $\Lambda\gtrsim1$, and the threshold in $\dot{\gamma}$ is smeared by the thermal nucleation.

\begin{figure*}[t]
\centering
\includegraphics[width=170mm]{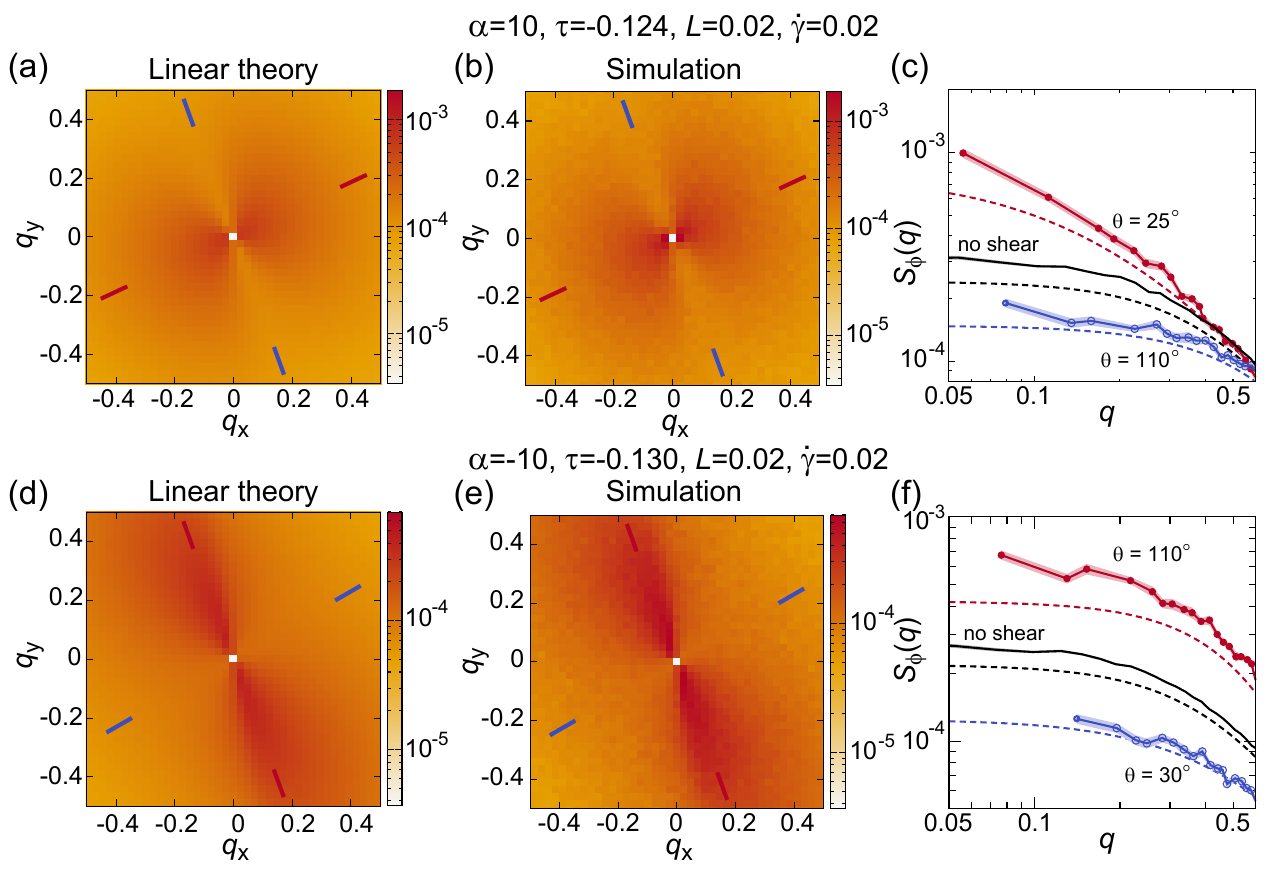}
\caption{
Structure factor of the order parameter $S_\phi(\bm q)$ of metastable liquid~1 under shear at $L=0.02$, $\dot{\gamma}=0.02$, and $V_0=0.01$. (a)--(c) $\alpha=10$ and $\tau=-0.124$. (d)--(f) $\alpha=-10$ and $\tau=-0.130$.
(a),(d) Linear theory, Eq.~(\ref{eq:Sk-steady}), and (b),(e) numerical simulation. The short lines indicate the directions used in (c) and (f). (c),(f) $S_\phi$ versus $q$ along the directions $\theta$ (angle of $\bm q$ from the $q_x$ axis) of the strongest enhancement (filled circles) and of the strongest suppression (open circles). The solid curves are the angular averages of $S_\phi$ without shear, and the shaded bands represent the standard errors. The dashed curves represent the linear theory, Eq.~(\ref{eq:Sk-steady}) for the two directions and Eq.~(\ref{eq:Sk-OZ}) without shear.
}
\label{fig::sk}
\end{figure*}

\subsubsection{Fluctuations in the metastable state}
\label{sec:sk-results}
Shear changes the fluctuations of liquid~1 also under the conditions where the transition does not occur, as explained in Sec.~\ref{sec:sk-theory}. Figure~\ref{fig::sk} shows the structure factor of the order parameter $S_\phi(\bm q)$ at $L=0.02$ and $\dot{\gamma}=0.02$ for $(\alpha, \tau)=(10, -0.124)$ and $(-10, -0.130)$, which correspond to $\dot{\gamma}/\dot{\gamma}_{\rm c}=1.0$ and $0.83$ and to $\Lambda=0.39$ and $0.36$, respectively. Under these conditions liquid~1 remains metastable, as shown in Fig.~\ref{fig::diagram}(b) and (d). $S_\phi(\bm q)$ obtained by the linear theory in Eq.~(\ref{eq:Sk-steady}) is presented in Fig.~\ref{fig::sk}(a) and (d) with the parameters of Sec.~\ref{sec:numparams}. They are compared with the numerical results shown in Fig.~\ref{fig::sk}(b) and (e). Qualitatively the same patterns are confirmed. For $\alpha=10$, $S_\phi$ is enhanced in the $q_xq_y>0$ direction and suppressed in the $q_xq_y<0$ direction, as shown in Fig.~\ref{fig::sk}(a-b). For $\alpha=-10$, the enhanced and suppressed directions are exchanged, as shown in Fig.~\ref{fig::sk}(d-e).

The angles $\theta$ at which the enhancement and the suppression are largest are also indicated by the red and blue short lines in Fig.~\ref{fig::sk} (a-b) and (d-e), where $\theta=25^\circ$ and $110^\circ$ for $\alpha=10$, and $\theta=110^\circ$ and $30^\circ$ for $\alpha=-10$, respectively. They deviate from $\theta=45^\circ$ and $135^\circ$. For $\alpha=10$, the enhancement is spread over a wide range of angles, as shown in Fig.~\ref{fig::sk}(b). For $\alpha=-10$, the enhancement is confined to a narrow angular range around $\theta=110^\circ$, as shown in Fig.~\ref{fig::sk}(e). Both the deviation and the difference between the two signs are due to the advection of the wavevector, as explained in Sec.~\ref{sec:sk-theory}.

The theoretical and numerical $S_\phi(\bm q)$ are compared quantitatively in Fig.~\ref{fig::sk}(c) and (f). The circles and the dashed curves represent the numerical and theoretical results, respectively. Though they are in good agreement, the linear theory underestimates the observed numerical values. Besides the contribution of the transverse velocity neglected in Sec.~\ref{sec:theory} (see Appendix~\ref{app:transverse}), the discrepancy arises from nonlinear fluctuations. For example, the average $\av{\phi}$ is displaced from its metastable value $\phi_{\rm e}$ as $\av{\phi}=-0.1284$ while $\phi_{\rm e}=-0.1300$ for $\alpha=10$ and $\tau=-0.124$. At the displaced average, $f_0''(\av{\phi})-K_0\alpha^2=0.330$, which is smaller than $\kappa'-K_0\alpha^2=0.415$ at $\phi_{\rm e}$ by approximately 20\%. The agreement between the theory and the simulation is improved when this nonlinear effect is incorporated. The nonlinear fluctuations also lower $\dot{\gamma}_{\rm c}$, which is proportional to $\kappa'-K_0\alpha^2$.

\begin{figure}[t]
\centering
\includegraphics[width=85mm]{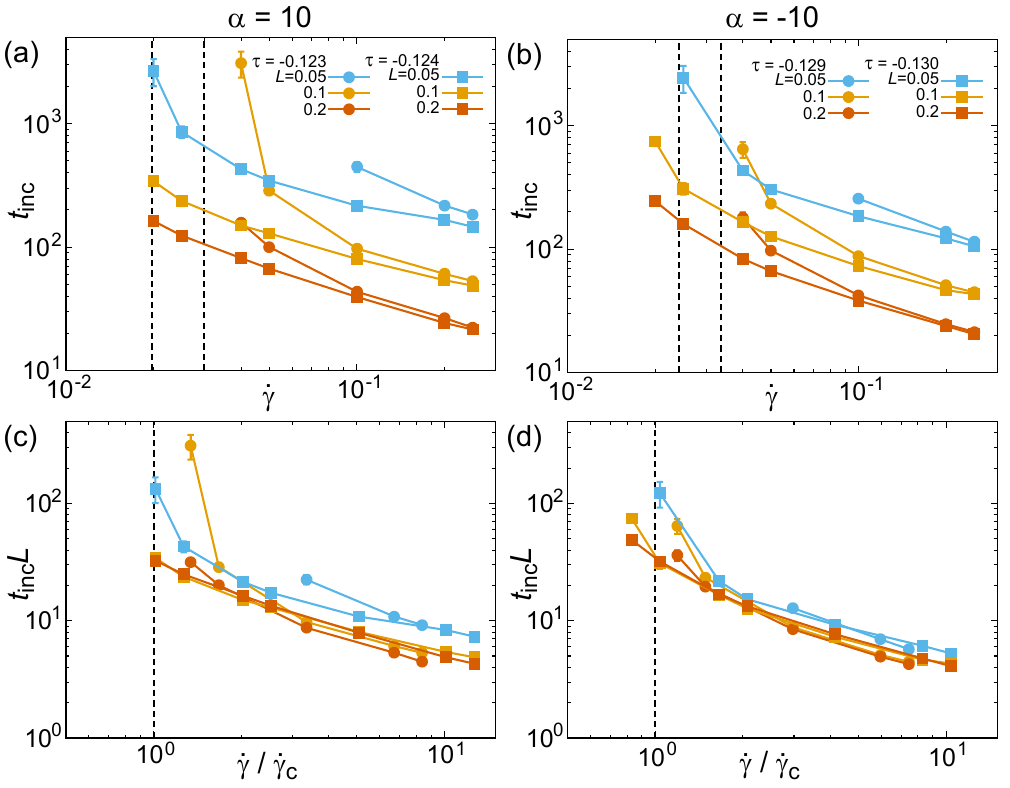}
\caption{
Incubation time $t_{\rm inc}$ for $V_0=0.01$. (a),(b) $t_{\rm inc}$ for (a) $\alpha=10$ and (b) $\alpha=-10$. (c),(d) The scaling plot $t_{\rm inc}L$ with respect to $\dot{\gamma}/\dot{\gamma}_{\rm c}$ for the data in (a) and (b), respectively. Circles and squares distinguish the two temperatures, and colors distinguish $L$. The vertical dashed lines indicate $\dot{\gamma}_{\rm c}$ at each temperature in (a) and (b), and $\dot{\gamma}=\dot{\gamma}_{\rm c}$ in (c) and (d). The symbols are the averages over six runs, and the error bars represent the standard errors.
}
\label{fig::scaling1}
\end{figure}

\begin{figure}[t]
\centering
\includegraphics[width=85mm]{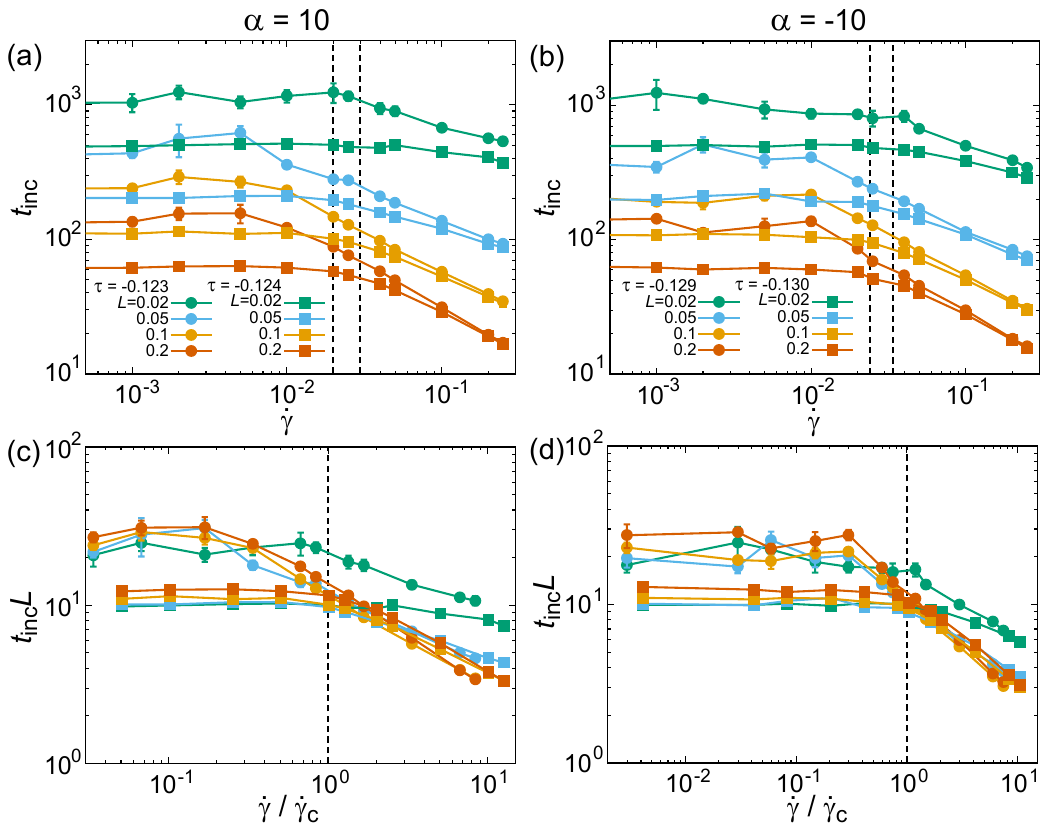}
\caption{
Incubation time $t_{\rm inc}$ for $V_0=0.02$, plotted as in Fig.~\ref{fig::scaling1}. The transition occurs at all the shear rates, including $\dot{\gamma}=0$.
}
\label{fig::scaling2}
\end{figure}

\begin{figure}[t]
\centering
\includegraphics[width=85mm]{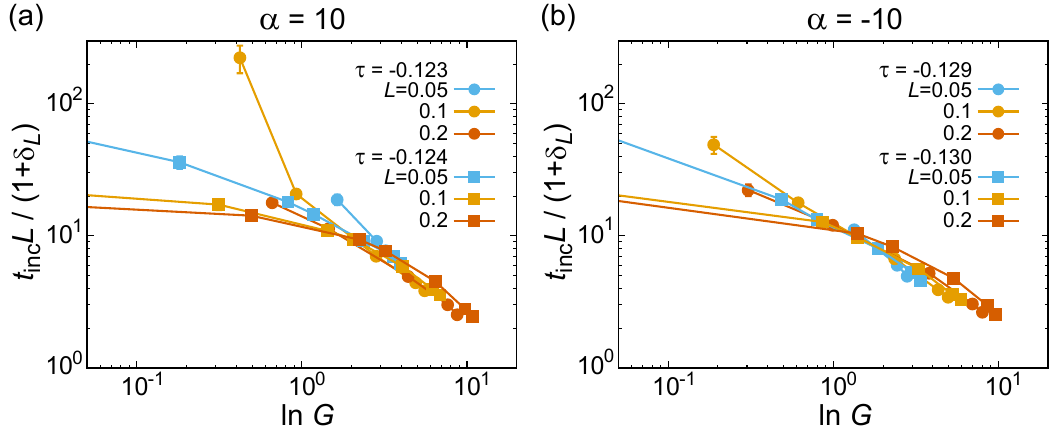}
\caption{
Incubation time for $V_0=0.01$ using the linear theory scaling for (a) $\alpha=10$ and (b) $\alpha=-10$. The incubation time $t_{\rm inc}$ is multiplied by $L/(1+\delta_L)$, to which the rate of the order-parameter mode is proportional in Eq.~(\ref{eq:appA-lameff0}). The horizontal axis is the amplification factor $\ln G=\Lambda\Phi(r)$ defined in Eq.~(\ref{eq:lnG}). The data are those of Fig.~\ref{fig::scaling1} for $\dot{\gamma}>\dot{\gamma}_{\rm c}$, and the symbols, the colors, and the error bars are the same as in Fig.~\ref{fig::scaling1}.
}
\label{fig::scaling3}
\end{figure}

\subsubsection{Acceleration of LLT under shear}
\label{sec:incubation}
When LLT is induced under shear, the incubation time depends on the shear rate. To examine the dependence, the incubation time for $V_0=0.01$ under the conditions where the transition occurs is presented in Fig.~\ref{fig::scaling1}. In (a) and (b), the incubation time decreases monotonically with increasing $\dot{\gamma}$. At large $\dot{\gamma}$, $t_{\rm inc}$ is nearly proportional to $1/L$ and depends only weakly on the temperature. In Fig.~\ref{fig::scaling1}(c) and (d), $t_{\rm inc}L$ is plotted against $\dot{\gamma}/\dot{\gamma}_{\rm c}$. The data for $L=0.1$ and $0.2$ at the two temperatures fall on a common curve for large $\dot{\gamma}/\dot{\gamma}_{\rm c}$, while the data for $L=0.05$ lie slightly above this curve. Near $\dot{\gamma}_{\rm c}$, the data deviate upward from the curve. For $\dot{\gamma}/\dot{\gamma}_{\rm c}\ge2$, $t_{\rm inc}L$ for $L=0.1$ and $0.2$ decreases as $(\dot{\gamma}/\dot{\gamma}_{\rm c})^{-\nu}$ with $\nu=0.70\pm0.05$. 
Although the variables $t_{\rm inc}L$ and $\dot{\gamma}/\dot{\gamma}_{\rm c}$ are convenient for displaying the data, this scaling is merely empirical and does not follow from the linear theory. In the linear theory, the rate of the order-parameter mode is proportional to $L/(1+\delta_L)$ and not to $L$, and the amplification depends not only on $\dot{\gamma}/\dot{\gamma}_{\rm c}$ but also on $\Lambda$, which contains $L$, as shown in Eq.~(\ref{eq:lnG}).

To compare the incubation time under shear with and without shear, the noise amplitude is increased to $V_0=0.02$. The result is presented in Fig.~\ref{fig::scaling2} (a) and (b). With this noise amplitude, the transition occurs by thermal nucleation without shear for $L\ge 0.02$. For $\dot{\gamma}\ll\dot{\gamma}_{\rm c}$, $t_{\rm inc}$ is nearly independent of $\dot{\gamma}$, and $t_{\rm inc}L$ depends only weakly on $L$, as shown in (c) and (d). The incubation time starts to decrease at $\dot{\gamma}\simeq\dot{\gamma}_{\rm c}$ for $(\alpha,\tau)=(10, -0.124)$ and $(-10, -0.130)$, whereas at $\dot{\gamma}\sim 0.2\dot{\gamma}_{\rm c}$ for $(10, -0.123)$ and $(-10,-0.129)$ when $L\ge0.05$. Shear thus accelerates the phase transition also below $\dot{\gamma}_{\rm c}$ due to the enhancement of the fluctuations. 
%For $\dot{\gamma}/\dot{\gamma}_{\rm c}\gtrsim2$, the difference between the two temperatures, which is a factor of 2--3 without shear, disappears, and the data for $L=0.1$ and $0.2$ approach a common curve as for $V_0=0.01$. The effect of shear is weaker for $L=0.02$: at $\tau=-0.124$ and $\dot{\gamma}=0.25$, $t_{\rm inc}$ is smaller than the value without shear by a factor of 2--4 for $L\ge0.05$ and by a factor of 1.3 for $L=0.02$. With the larger noise, shear thus changes the rate of a nucleation that occurs also without shear, and no boundary appears in the $(\dot{\gamma},L)$ plane.

To relate the incubation time at high shear rates to the linear theory, $t_{\rm inc}$ is expressed with the rate of the order-parameter mode, which is proportional to $L/(1+\delta_L)$ from Eq.~(\ref{eq:appA-lameff0}), and the amplification factor $\ln G=\Lambda\Phi(r)$ in Eq.~(\ref{eq:lnG}). In Fig.~\ref{fig::scaling3}, $t_{\rm inc}L/(1+\delta_L)$ for $V_0=0.01$ and $\dot{\gamma}>\dot{\gamma}_{\rm c}$ is plotted against $\ln G$. For $\ln G\gtrsim1$, the data collapse to a master curve. On the master curve, $t_{\rm inc}L/(1+\delta_L)$ decreases as $(\ln G)^{-\nu_G}$ with $\nu_G=0.80\pm0.06$. 
The master curve has two implications. First, the temperature enters $\ln G$ only through $\dot{\gamma}_{\rm c}$, because $\Lambda$ is independent of the temperature: both $|\lambda_0|$ and $\dot{\gamma}_{\rm c}$ are proportional to $\kappa'-K_0\alpha^2$, and $\delta_L$ contains $\kappa'$, whose temperature dependence is small. The collapse of the data at the two temperatures onto the master curve therefore shows that the temperature changes the incubation time only through $\dot{\gamma}_{\rm c}$. Second, the scaling of $t_{\rm inc}L$ in Fig.~\ref{fig::scaling1} results from a cancellation. At a fixed $\dot{\gamma}/\dot{\gamma}_{\rm c}$, an increase of $L$ increases $\ln G$, which lowers $t_{\rm inc}L/(1+\delta_L)$, and increases the factor $1+\delta_L$, so that $t_{\rm inc}L\propto(1+\delta_L)^{1+\nu_G}L^{-\nu_G}$ changes only slightly.

The scaling in Fig.~\ref{fig::scaling3} does not hold for $\ln G<1$, where $t_{\rm inc}$ depends on the noise amplitude. The ratio of $t_{\rm inc}$ for $V_0=0.01$ to that for $V_0=0.02$ under the same conditions is 2--30 for $\ln G<1$. Close to $\dot{\gamma}_{\rm c}$, the transition is thus triggered by a rare nucleation event, as in the case without shear, whereas for large $\ln G$ the incubation time is determined by the amplification under shear flow.

\begin{figure}[t]
\centering
\includegraphics[width=85mm]{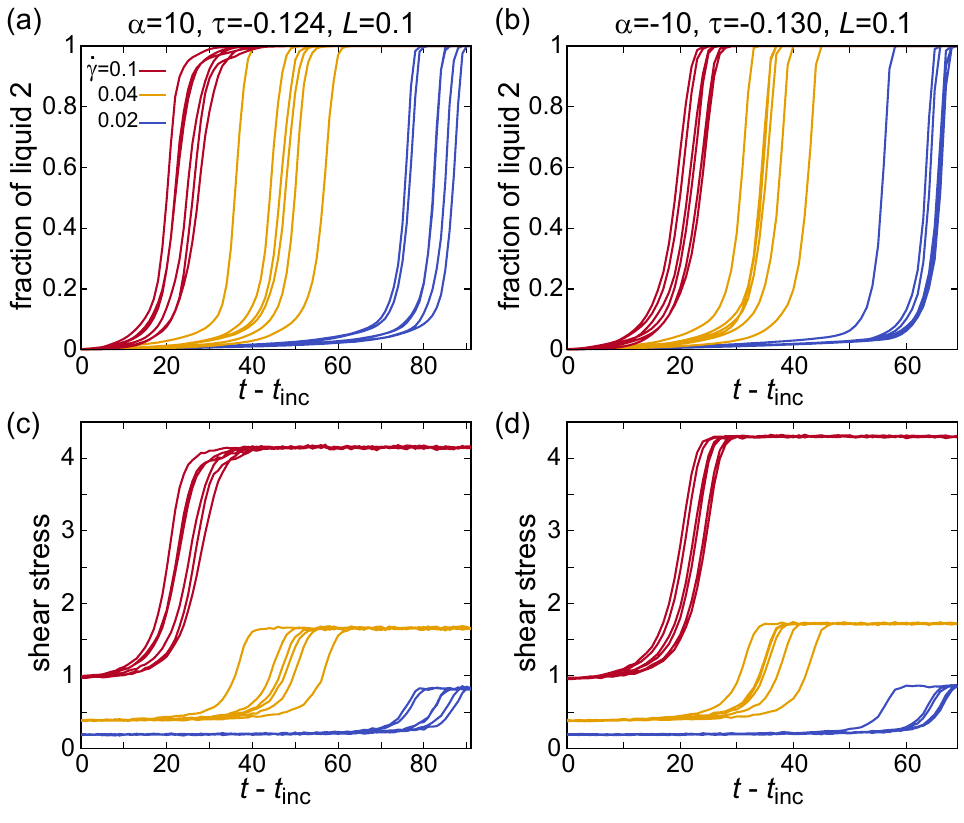}
\caption{
(a),(b) Fraction of liquid~2 and (c),(d) spatial average of the shear stress $\sigma_{xy}$ as functions of $t-t_{\rm inc}$ for $L=0.1$ and $V_0=0.01$. (a),(c) $\alpha=10$ and $\tau=-0.124$. (b),(d) $\alpha=-10$ and $\tau=-0.130$. The shear rates are $\dot{\gamma}=0.02$, $0.04$, and $0.1$. Each curve represents a single run.
}
\label{fig::growth}
\end{figure}

\begin{figure}[t]
\centering
\includegraphics[width=85mm]{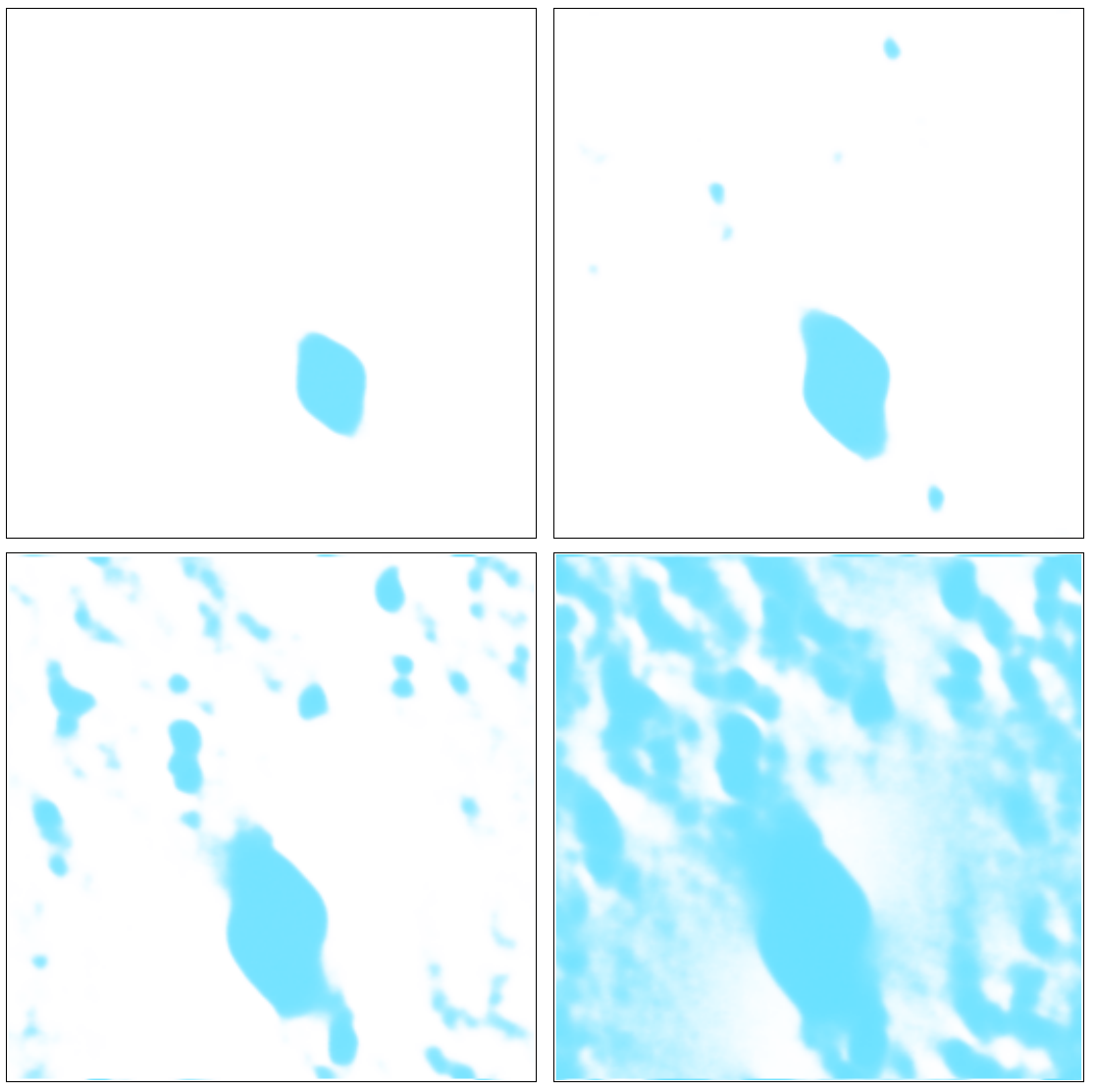}
\caption{
Patterns of $\phi$ after nucleation for $\alpha=10$, $\tau=-0.124$, $L=0.1$, $\dot{\gamma}=0.02$, and $V_0=0.01$. In this run $t_{\rm inc}=384$. Time increases from the top left to the bottom right panel: $t-t_{\rm inc}=56$, $71$, $81$, and $86$. White and blue regions represent liquid~1 and liquid~2, respectively. The flow is in the horizontal ($x$) direction, and the velocity gradient is in the vertical ($y$) direction.
}
\label{fig::snapshot1}
\end{figure}

\begin{figure}[t]
\centering
\includegraphics[width=85mm]{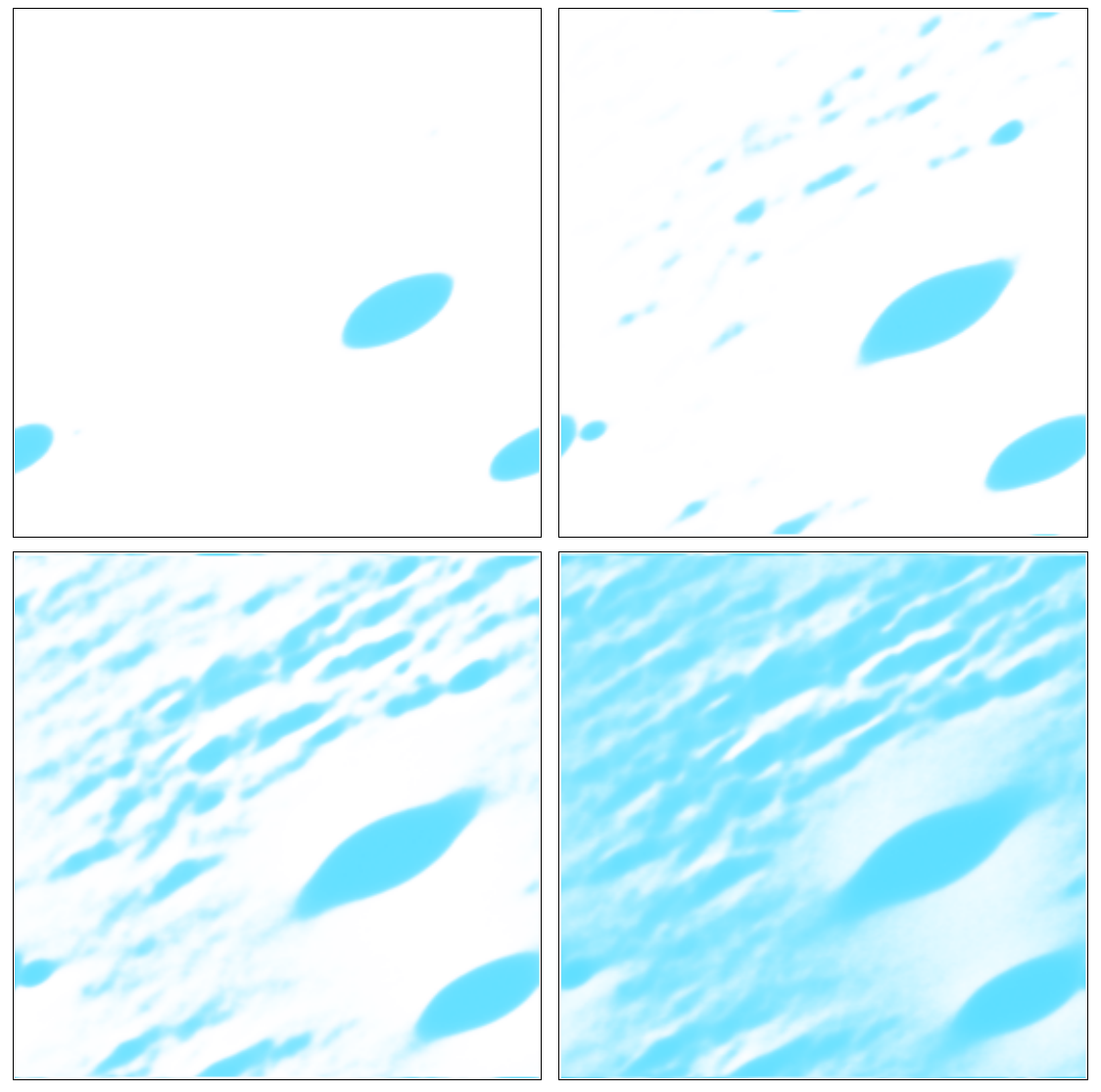}
\caption{
Patterns of $\phi$ after nucleation for $\alpha=-10$, $\tau=-0.130$, $L=0.1$, $\dot{\gamma}=0.02$, and $V_0=0.01$, shown as in Fig.~\ref{fig::snapshot1}. In this run $t_{\rm inc}=674$. The times are $t-t_{\rm inc}=41$, $51$, $55$, and $57$.
}
\label{fig::snapshot2}
\end{figure}

\subsubsection{Growth after nucleation}
The transformation after nucleation proceeds in two stages. The fraction of liquid~2 first increases slowly and then rises steeply to unity, as shown in Fig.~\ref{fig::growth}(a) and (b). The completion time decreases with increasing $\dot{\gamma}$. In the figure it is $\sim 80$, 50, and 20 for $\dot{\gamma}=0.02$, $0.04$, and $0.1$ at $\alpha=10$, and $\sim 60$, 40, and 20 at $\alpha=-10$, while the time interval of the steep rise is 5--10 at all the shear rates. Thus the shear flow mainly changes the first stage. For $\dot{\gamma}\gtrsim2\dot{\gamma}_{\rm c}$, the completion time decreases with $\dot{\gamma}$ in the same way as the incubation time. This suggests that both are governed by the same kinetics of the order parameter enhanced by shear. As $\dot{\gamma}$ decreases toward $\dot{\gamma}_{\rm c}$, the incubation time increases more steeply than the completion time, because the incubation near $\dot{\gamma}_{\rm c}$ is limited by the thermal nucleation (Sec.~\ref{sec:incubation}) whereas the growth is not.

Here it should be noted that the acceleration of the growth differs in origin from that in systems with a conserved order parameter, where flow accelerates the growth of droplets in a binary mixture~\cite{baumberger1992kinetics}, of dendrites growing against the flow~\cite{tong2001phase}, and of colloidal crystals~\cite{lander2013crystallization} by transporting the solute, the heat, or the particles to the interface. In the LLT, the order parameter is non-conserved, so that the growth does not require the diffusive transport of $\phi$ to the interface. The density difference between the two liquids is accommodated by the compressional flow~\cite{takae2020role}, not by diffusion.

The origin of the two stages is evident in the patterns of $\phi$ after nucleation, as shown in Figs.~\ref{fig::snapshot1} ($\alpha=10$) and \ref{fig::snapshot2} ($\alpha=-10$). Isolated domains of liquid~2 first appear and grow. Subsequently, further domains with the same anisotropic morphology appear in the region away from the first domains. In the last stage, $\phi$ increases over the remaining region, and the whole system becomes liquid~2. The slow increase of the fraction in Fig.~\ref{fig::growth}(a) and (b) thus corresponds to the growth of the first domains, and the steep rise corresponds to the appearance of the further domains and the transformation of the remaining region.

The shape of the first domains shows how shear selects the morphology of the nucleus, which deviates from an isotropic shape. They are elongated in the $xy<0$ direction for $\alpha=10$ and in the $xy>0$ direction for $\alpha=-10$. These directions are perpendicular to the wavevectors of the enhanced fluctuations in Fig.~\ref{fig::sk}. For $\alpha=10$, the elongation is along the compression axis of the shear flow, which is opposite to the deformation of a passively advected domain. The shape of the nucleus in the early stage thus reflects the anisotropic growth of the order parameter.
The flow also deforms the domains in the late stage, and the deformation depends on the sign of $\alpha$. A passively advected domain is stretched along the extension axis $xy>0$. For $\alpha=10$, this axis differs from the elongation axis of the nucleus, so that the advection reduces the elongation, as shown in Fig.~\ref{fig::snapshot1}. For $\alpha=-10$, the two axes coincide, so that the advection increases the elongation, as shown in Fig.~\ref{fig::snapshot2}. The same rotation of the wavevector underlies the anisotropy of the structure factor in Fig.~\ref{fig::sk}. The enhanced wavevectors are rotated toward the $q_x$ axis for $\alpha=10$ and toward the $q_y$ axis for $\alpha=-10$ (Sec.~\ref{sec:sk-theory}), which in real space corresponds to the reduction and the increase of the elongation, respectively. The narrower enhanced region in $S_\phi$ and the stronger elongation of the domains for $\alpha=-10$ thus have a common origin.

Because the two liquids differ in viscosity, the transition changes the shear stress. The spatial average of the shear stress $\sigma_{xy}$ is displayed in Fig.~\ref{fig::growth}(c) and (d).
%Because the two liquids differ in viscosity, the transition changes the shear stress. The spatial average of the viscous shear stress $\sigma_{xy}$ is displayed in Fig.~\ref{fig::growth}(c) and (d). The reversible stress $\Pi_{xy}=D\nabla_x\phi\nabla_y\phi$ from the interfaces is not included; it vanishes in the homogeneous liquids before and after the transition and contributes only during the transition. 
Before the transition, $\sigma_{xy}$ agrees with $\eta(\phi_{\rm e})\dot{\gamma}$ of homogeneous liquid~1. After the transition, $\sigma_{xy}$ agrees with $\eta\dot{\gamma}$ of homogeneous liquid~2 at the same temperature and pressure. The transition thus increases the apparent viscosity $\sigma_{xy}/\dot{\gamma}$. The stress changes little during the slow increase of the fraction and increases together with the steep rise. A measurement of the viscosity under shear therefore detects the transition with a delay equal to the duration of the first stage, 20--80 ($\lesssim 0.2 t_{\rm inc}$) in Fig.~\ref{fig::growth}.

\section{Discussion and summary}
\label{sec:discussion}

%Furukawa
This study reveals that the effect of shear on an LLT is not determined only by the instability condition at a fixed wavevector, $\dot{\gamma}>\dot{\gamma}_{\rm c}$. Because the wavevector of a fluctuation is advected by the flow, the fluctuation grows only for a limited time duration, as known in the stability theory of shear flows~\cite{schmid2007nonmodal}, and the amplification depends on a second number, $\Lambda=|\lambda_0|/\dot{\gamma}_{\rm c}$. The second number distinguishes the LLT from the shear-induced instability of one-component liquids and glasses~\cite{furukawa2006violation,furukawa2009inhomogeneous}, in which the viscosity depends on the conserved density and the instability occurs for wavevectors in particular directions above a critical shear rate determined by material parameters. In the density-coupled case, the counterpart of $\Lambda$ is $(\dot{\gamma}_{\rm c}\tau_\rho)^{-1}\sim\p\ln\eta/\p\ln\rho >1$, where $\tau_\rho$ is the relaxation time of the density fluctuation. Then the condition $\dot{\gamma}>\dot{\gamma}_{\rm c}$ obtained at a fixed wavevector is sufficient. In the current model of the LLT, $|\lambda_0|$ is proportional to the kinetic coefficient $L$ of the order parameter, and $\dot{\gamma}_{\rm c}$ is determined by the viscosity coupling $\eta_\psi$. These are distinct material parameters, and $\Lambda\propto L\eta_\psi$ can be smaller or larger than unity. The condition $\dot{\gamma}>\dot{\gamma}_{\rm c}$ is thus not sufficient for the shear-induced LLT, and $\Lambda\gtrsim1$ is also required, as the absence of the transition at $\Lambda\simeq0.4$ in Fig.~\ref{fig::diagram} shows.

%experiment
For real LLT systems, $\Lambda$ may be estimated by assuming that the kinetic coefficient scales as $L\propto\tau_\alpha^{-1}$, where $\tau_\alpha$ is the molecular structural relaxation time~\cite{murata2019link}. If the viscosity also scales as $\eta\propto\tau_\alpha$, as expected from a Maxwell-type relation~\cite{Onukibook,HansenMcDonald}, the product $L\eta$ becomes weakly dependent on $\tau_\alpha$. Writing $\eta_\psi=\eta a$ with $a=\p\ln\eta/\p\psi$ on the liquid-1 side,
\be
\Lambda\sim K_0'|\alpha|(L\eta)a,
\label{eq:Lambda-exp}
\en
so that, under these assumptions, $\Lambda$ depends primarily on the logarithmic viscosity derivative $a$ rather than on the absolute viscosity.
Although the viscosity contrast between the two liquids reported for TPP-type~\cite{tanaka2004liquid,kurita2006kinetics,kurita2005control} and water-type~\cite{gallo2016water,shi2018origin} systems is much larger than that in the present simulations, the linear mechanism is controlled by $a$ and not by the total contrast. The local slope can be determined from the fraction of locally favored structures, which is accessible experimentally~\cite{murata2015microscopic} or from two-state fits~\cite{caupin2019thermodynamics}, combined with the viscosity~\cite{dehaoui2015viscosity}. The critical shear rate is $\dot{\gamma}_{\rm c}=|\lambda_0|/\Lambda$ with $|\lambda_0|\sim L(\kappa'-K_0\alpha^2)\sim(\kappa'-K_0\alpha^2)/\tau_\alpha$, so that $\dot{\gamma}_{\rm c}\tau_\alpha\sim(\kappa'-K_0\alpha^2)/\Lambda$. Thus, sufficiently close to the spinodal, where $\kappa'-K_0\alpha^2$ becomes small, the critical shear rate can become smaller than $\tau_\alpha^{-1}$.

%metallic glass
The only LLT experiment under controlled shear is the bulk metallic glass former Zr$_{41.2}$\allowbreak Ti$_{13.8}$\allowbreak Cu$_{12.5}$\allowbreak Ni$_{10}$\allowbreak Be$_{22.5}$~\cite{way2007influence,gallino2021effect}. When the fragile, low-viscosity liquid of this system is supercooled and sheared, the viscosity remains constant during an incubation period and then rises toward that of the strong liquid, and the onset time of the rise decreases with the shear rate approximately as $\dot{\gamma}^{-0.6}$~\cite{gallino2021effect}. The acceleration was described with the shear-modified classical nucleation theory~\cite{mura2016effects}, in which shear enhances the transport of particles to the nucleus~\cite{batchelor1979mass}. The viscosity coupling $\eta_\psi$ provides another mechanism, which does not involve the transport of particles to the nucleus, because the structural order parameter is non-conserved. In the simulations, the incubation time for $\dot{\gamma}\ge2\dot{\gamma}_{\rm c}$ and $\Lambda\gtrsim1$ decreases as $\dot{\gamma}^{-\nu}$ with $\nu=0.70\pm0.05$ in Fig.~\ref{fig::scaling1}, and the growth of the viscosity follows, as shown in Fig.~\ref{fig::growth}. The dependence on the shear rate is thus comparable. If $\Lambda\ll1$, the acceleration by shear should be weak. If the liquid is supercooled toward the spinodal of the LLT, $\dot{\gamma}_{\rm c}$ decreases, and the acceleration should be observed at lower shear rates.

%summary
In summary, the dependence of the viscosity on the non-conserved structural order parameter leads to the occurrence and acceleration of the LLT from metastable liquid under shear. Shear changes the susceptibility of the structural order parameter fluctuation depending on the direction of its wavevector, and the flow rotates the wavevector. The effect of shear is therefore determined by $\dot{\gamma}/\dot{\gamma}_{\rm c}$ and $\Lambda$, the latter of which measures how rapidly the amplitude of a fluctuation follows the rotation. In the hydrodynamic simulations, the fluctuations of the metastable liquid are enhanced in the directions given by the linear theory, though incorporating nonlinear fluctuations is needed to confirm quantitative agreement. The transition occurs for $\dot{\gamma}\gtrsim\dot{\gamma}_{\rm c}$ and $\Lambda\gtrsim1$ by nucleation of domains with anisotropic morphology. The shear-rate dependence of the incubation time gives estimates of $\dot{\gamma}_{\rm c}$ and $\Lambda$, and the elongation direction of the growing domains gives the sign of $\alpha\eta_\psi$.
%The two conditions do not define a sharp threshold, because the nucleation is activated by the thermal noise. Far above $\dot{\gamma}_{\rm c}$, the incubation time multiplied by $L/(1+\delta_L)$ is a function of the amplification factor $G$ of the linear theory. The form of this function and the incubation time near $\dot{\gamma}_{\rm c}$, where the barrier for nucleation under shear enters, require a treatment of the nonlinear effects. Because the absolute viscosity cancels from $\Lambda$, the criterion applies to molecular liquids and metallic melts near their glass transition, and the shear-rate dependence of the onset time of the LLT gives a way to estimate $\dot{\gamma}_{\rm c}$ and $\Lambda$ experimentally. The later stage of the transformation for the much larger viscosity contrast of real systems remains open.

\begin{acknowledgments}
This study was supported by JSPS Kakenhi Grant Nos.~JP24K00594, JP25H01978, and JP26H00386.
\end{acknowledgments}

\section*{AUTHOR DECLARATIONS}
\subsection*{Conflict of Interest}
The author declares no competing interests.

\subsection*{Author Contributions}
\textbf{Kyohei Takae}: Conceptualization (lead); Methodology (lead); Software (lead); Formal analysis (lead); Investigation (lead); Visualization (lead); Writing – original draft (lead); Writing – review and editing (lead).

\section*{DATA AVAILABILITY}
The data that supports the findings of this study are available from the corresponding author upon reasonable request.

\appendix
\makeatletter
\@addtoreset{equation}{section}
\makeatother
\renewcommand{\theequation}{\thesection.\arabic{equation}}

\section{Effect of the transverse velocity}
\label{app:transverse}

The linear theory of Sec.~\ref{sec:theory} neglects the first term on the r.h.s. of Eq.~(\ref{eq:Z}), $-2\rho_{\rm e}\dot{\gamma}\nabla_xV_y$, which couples the shear flow to the transverse velocity. The transverse momentum diffuses at the rate $\eta_{\rm e}q^2/\rho_{\rm e}$, so that the neglect requires $\delta_\perp=2\rho_{\rm e}\dot{\gamma}/\eta_{\rm e}q^2\ll1$. For the parameters of Sec.~\ref{sec:numparams}, $\delta_\perp\simeq 1$ at $q=0.1$ for $\dot{\gamma}=0.05$ and at $q=0.063$ for $\dot{\gamma}=0.02$, which lies inside the range of $q$ compared with the simulations. To examine the effect of this term, the linearized equations (\ref{eq:lm})--(\ref{eq:lv}) for $(m,\psi,V_x,V_y)$ are solved numerically with the parameters of Sec.~\ref{sec:numparams}.

\begin{table}[t]
\caption{
Growth rate $\lambda_\psi$ of the order-parameter mode
at a fixed wavevector. The values are obtained from the real eigenvalue associated with this mode in Eq.~(\ref{eq:matrix}) ($3\times3$) and in the full linearized equations (full). The parameters are $L=0.1$ and $\dot{\gamma}=0.05$, with $\tau=-0.124$ for $\alpha=10$ and $\tau=-0.130$ for $\alpha=-10$. The wavevector is oriented along the direction of strongest growth, $\theta=\pi/4$ for $\alpha=10$ and $\theta=3\pi/4$ for $\alpha=-10$.}
\label{tab:lam}
\begin{ruledtabular}
\begin{tabular}{ccccc}
$\alpha$ & $q$ & $3\times3$ & full & difference \\
\hline
$10$  & 0.05 & 0.0372 & 0.0315 & $-15\%$ \\
$10$  & 0.10 & 0.0414 & 0.0399 & $-4\%$  \\
$10$  & 0.20 & 0.0390 & 0.0387 & $-1\%$  \\
$-10$ & 0.05 & 0.0314 & 0.0337 & $+7\%$  \\
$-10$ & 0.10 & 0.0340 & 0.0349 & $+3\%$  \\
$-10$ & 0.20 & 0.0307 & 0.0310 & $+1\%$  \\
\end{tabular}
\end{ruledtabular}
\end{table}

\begin{table}[t]
\caption{Structure factor $S_\phi(\bm q)$ in units of $10^{-4}$ in the direction of the strongest enhancement under the conditions of Fig.~\ref{fig::sk} ($L=0.02$, $\dot{\gamma}=0.02$, $V_0=0.01$), obtained from Eq.~(\ref{eq:appA-lyap}) for the $3\times3$ and the full linearized equations. The directions are $\theta=25^\circ$ for $\alpha=10$ and $110^\circ$ for $\alpha=-10$. The $3\times3$ values agree with Eq.~(\ref{eq:Sk-steady}).}
\label{tab:sk}
\begin{ruledtabular}
\begin{tabular}{ccccc}
$\alpha$ & $q$ & $3\times3$ & full & difference \\
\hline
$10$  & 0.05 & 6.34 & 6.77 & $+7\%$ \\
$10$  & 0.10 & 4.92 & 4.89 & $-1\%$ \\
$10$  & 0.20 & 3.11 & 3.09 & $-1\%$ \\
$-10$ & 0.05 & 3.94 & 9.19 & $+133\%$ \\
$-10$ & 0.10 & 3.99 & 4.54 & $+14\%$ \\
$-10$ & 0.20 & 3.69 & 3.94 & $+7\%$ \\
\end{tabular}
\end{ruledtabular}
\end{table}

Table~\ref{tab:lam} compares the growth rate at a fixed wavevector. In the direction of the strongest growth, the transverse velocity changes $\lambda_\psi$ by $-15\%$ ($\alpha=10$) and $+7\%$ ($\alpha=-10$) at $q=0.05$, where $\delta_\perp\simeq4$, and by less than 4\% for $q\ge0.1$. The sign of the change is opposite for the two signs of $\alpha$, because the coupling reduces the growth for $|q_y/q_x|<1$ and increases it for $|q_y/q_x|>1$. In the directions of suppression, the full linearized equations have in addition a slowly decaying mode of the transverse velocity at $q\lesssim0.1$, whose coupling with $\psi$ is weak. 
Table~\ref{tab:sk} shows the results for the structure factor, which is computed as follows. Writing $V_{x,\bm q}=ia$ and $V_{y,\bm q}=ib$ with real $a$ and $b$, the linearized equations (\ref{eq:lm})--(\ref{eq:lv}) for $\bm x=(m,\psi,a,b)$ take the form $\dot{\bm x}=A(\bm q)\bm x+\bm\xi$ with real coefficients, where $A$ is the coefficient matrix and $\bm\xi=(0,\theta_{\bm q},\xi_a,\xi_b)$ is the noise term. Along the trajectory (\ref{eq:appA-lagq}), the covariance matrix $C=\av{\bm x\bm x^{\rm T}}$ obeys
\be
\dot{C}=AC+CA^{\rm T}+N,
\label{eq:appA-lyap}
\en
where $N$ denotes the noise matrix $\av{\xi_i(t)\xi_j(t')}=N_{ij}\delta(t-t')$.
%From Eqs.~(\ref{eq:fluc1}) and (\ref{eq:fluc2}), its nonzero elements are $N_{\psi\psi}=2LV_0^2$ and
%\be
%N_{aa}=\frac{2k_{\rm B}TV_0^2}{\rho_{\rm e}^2}\big[\eta_{\rm e}q^2+(\zeta_{\rm e}-\tfrac{2}{d}\eta_{\rm e}+\eta_{\rm e})q_x^2\big],
%\quad
%N_{ab}=\frac{2k_{\rm B}TV_0^2}{\rho_{\rm e}^2}(\zeta_{\rm e}-\tfrac{2}{d}\eta_{\rm e}+\eta_{\rm e})q_xq_y,
%\en
%and $N_{bb}$ with $q_x^2$ replaced by $q_y^2$ in $N_{aa}$, where $V_0^2=v_0/\ell_0^d$ converts the noise amplitudes to the dimensionless units. 
The steady state is obtained by integrating Eq.~(\ref{eq:appA-lyap})
%from $\mathsf C=0$ far upstream on the trajectory until the solution becomes independent of the starting point, 
and $S_\phi(\bm q)=C_{\psi\psi}$. The $3\times3$ values are obtained in the same way from Eq.~(\ref{eq:matrix}) with $\bm x=(m,\psi,Z)$.
%, $N_{\psi\psi}=2LV_0^2$, and $N_{ZZ}=2k_{\rm B}TV_0^2\eta_{\rm e}'q^4/\rho_{\rm e}^2$.
%Without shear, Eq.~(\ref{eq:appA-lyap}) is solved directly for $\dot{\mathsf C}=0$ and reproduces Eq.~(\ref{eq:Sk-OZ}) within 1\%.
For $\alpha=10$, $S_\phi$ changes by 7\% at $q=0.05$ and by 1\% for $q\ge0.1$. For $\alpha=-10$, 
%a fluctuation reaching the enhanced direction has passed through the perpendicular direction, where it is no longer suppressed at small $q$;
$S_\phi$ is larger than the $3\times 3$ values by 14\% at $q=0.1$ and 133\% at $q=0.05$. Thus the difference between the theory and the simulation at $q\le0.1$ in Fig.~\ref{fig::sk}(f) also arises from the transverse velocity. The neglect of the transverse velocity is thus justified for $q\gtrsim 0.05$ for $\alpha=10$ and $q\gtrsim 0.1$ for $\alpha=-10$ in the present numerical parameters.

\section{Isobaric reference state and spinodal of the simulated system}
\label{app:isobaric}

The coefficients of Eq.~(\ref{eq:gammac}), i.e., $\kappa'$, $m_{\rm e}$, $\eta_\psi$, must be evaluated at the homogeneous state at which the hydrodynamic equations are linearized. In the isobaric ensemble, this state is evaluated under the two conditions
\be
\Gamma=f_0'(\phi_{\rm e})-\alpha m_{\rm e}=0,
\qquad
p(\phi_{\rm e},m_{\rm e})=P_0.
\label{eq:appD-state}
\en
The second condition yields
\be
m_{\rm e}=-1+\sqrt{1+2K_0\big[P_0-\mu_0n_0+f_0(\phi_{\rm e})+\alpha\phi_{\rm e}\big]}.
\label{eq:appD-miso}
\en
This must be distinguished from the elimination $\p f/\p m=0$ (i.e., $m=K_0\alpha\phi$), which yields the renormalized potential with $\kappa\to\kappa-K_0\alpha^2$. Along the isobar, differentiating $p=P_0$ gives $dm/d\phi=K_0(f_0'+\alpha)/(1+m)$, which reduces to $dm/d\phi=K_0\alpha$ at $\Gamma=0$. Thus $d\Gamma/d\phi=f_0''-K_0\alpha^2$ needs to be evaluated at the isobaric state $\phi_{\rm e}(\tau,P_0)$.
%Hence $d\Gamma/d\phi=f_0''-K_0\alpha^2$ and the spinodal criterion of the main text, $\kappa'=K_0\alpha^2$, is unchanged, but it must be imposed on the isobar $\phi_{\rm e}(\tau;P_0)$.
For the simulation parameters, the resulting spinodal reads
\begin{align}
\tau_{\rm s}(P_0=0,\alpha=10)&=-0.1248,\\
\tau_{\rm s}(P_0=0,\alpha=-10)&=-0.1311.
\label{eq:appD-taus}
\end{align}

\section{Finite-$L$ correction to the growth rate}
\label{app:deltaL}

The growth rate $\lambda_\psi$ of Sec.~\ref{sec:advection} was obtained to the lowest order in $L$, where the density follows the order parameter instantaneously. Here we derive the leading correction in $L$. It is used in the values of $\Lambda$ and $\lambda_0$ quoted in Sec.~\ref{sec:numerical}. Eliminating $Z$ from Eq.~(\ref{eq:matrix}) and substituting into $\dot{m}=-(1+m_{\rm e})Z$ yields
\be
\dot m=-\Gamma_\rho\big[m-K_0'(\alpha'+\dot{\gamma}w\eta_\psi)\psi\big],
\label{eq:appC-mdot}
\en
where $\Gamma_\rho=(1+m_{\rm e})/K_0'\eta_{\rm e}'$ is the rate at which the density relaxes against the viscous stress ($\Gamma_\rho\simeq1.2$ for the parameters of Sec.~\ref{sec:numparams}). By setting $m,\psi\propto e^{\lambda t}$, Eq.~(\ref{eq:appC-mdot}) gives $m=K_0'(\alpha'+\dot{\gamma}w\eta_\psi)\psi/(1+\lambda/\Gamma_\rho)$. Substituting this into $\dot\psi=L\alpha m-L(\kappa'+Dq^2)\psi$ yields
\be
\lambda\Big[1+\delta_L+\frac{\lambda}{\Gamma_\rho}\Big]=\lambda_\psi^{(0)},
\label{eq:appC-quad}
\en
where $\delta_L=L\kappa'/\Gamma_\rho$ is the ratio of the relaxation rate of $\psi$ at a fixed density to that of the density with $Dq^2$ being neglected. $\lambda_\psi^{(0)}=L[\alpha K_0'(\alpha'+\dot{\gamma}w\eta_\psi)-\kappa'-Dq^2]$ is given in Eq.~(\ref{eq:appA-lameff0}). The term $\lambda/\Gamma_\rho$ is dropped. Retaining it makes $\lambda_\psi$ a nonlinear function of $w-r$ and breaks the form of Eqs.~(\ref{eq:appA-lameff})--(\ref{eq:lnG}) and (\ref{eq:Sk-long}), whereas it vanishes at $\dot{\gamma}_{\rm c}$ and does not change the critical shear rate. Its effect, together with that of the rotation of the wavevector at the rate $\dot{\gamma}$, is evaluated by integrating Eq.~(\ref{eq:matrix}) numerically along the trajectory (\ref{eq:appA-lagq}): for $L=0.2$ and $\dot{\gamma}\ge5\dot{\gamma}_{\rm c}$, the omitted terms reduce $\ln G$ by 10--25\% below the integral of $\lambda_\psi$ given below, and for $\dot{\gamma}\le2\dot{\gamma}_{\rm c}$ by less than 5\%. Dropping the term gives
\be
\lambda_\psi=\frac{\lambda_\psi^{(0)}}{1+\delta_L}.
\label{eq:appC-lam}
\en
The dimensionless number $\Lambda$ in Eq.~(\ref{eq:appA-Lambda}) is then modified as
\be
\Lambda=\frac{|\lambda_0|}{(1+\delta_L)\dot{\gamma}_{\rm c}}.
\label{eq:appC-Lambda}
\en
The values of $\delta_L/L$ in Sec.~\ref{sec:numparams} are $\delta_L/L\sim 3$, so that $\delta_L$ ranges from $0.06$--$0.08$ at $L=0.02$ to $0.6$--$0.8$ at $L=0.2$.

%Retaining the term $\lambda/\Gamma_\rho$ in Eq.~(\ref{eq:appC-quad}) gives the closed form $\lambda_\psi=(\Gamma_\rho/2)[-(1+\delta_L)+\sqrt{(1+\delta_L)^2+4\lambda_\psi^{(0)}/\Gamma_\rho}\,]$, which differs from Eq.~(\ref{eq:appC-lam}) by the factor $1/[1+\lambda_\psi/\Gamma_\rho(1+\delta_L)]$. The difference is at most $10\%$ for $L=0.2$ and $\dot{\gamma}=5\dot{\gamma}_{\rm c}$, and Eq.~(\ref{eq:appC-lam}) is used because it keeps $\lambda_\psi$ linear in $w-r$, on which Eqs.~(\ref{eq:appA-lameff})--(\ref{eq:lnG}) and (\ref{eq:Sk-steady}) rely.

%Without the overdamped approximation, the slow root of the characteristic equation (\ref{eq:chara}) satisfies
%\be
%\lambda\Big[1+\delta_L+\frac{\lambda}{\Gamma_\rho}+\frac{\lambda\,(\lambda+L\kappa')}{c_{\rm s}^2q^2}\Big]=\lambda_\psi^{(0)},
%\label{eq:appC-exact}
%\en
%where the last term is the inertial correction neglected in Eq.~(\ref{eq:appC-mdot}). It becomes comparable to $\delta_L$ only for $q^2\eta_{\rm e}'/\rho_{\rm e}\lesssim|\lambda_\psi|$, i.e., at the smallest wavenumbers for $L=0.2$, where the density responds inertially and the lag is weaker than Eq.~(\ref{eq:appC-lam}) gives. Equation~(\ref{eq:appC-lam}) differs from the root of Eq.~(\ref{eq:appC-exact}) by less than $1.1\%$ for $L=0.02$ and by up to $7\%$ for $L=0.2$ at $q=0.6$ in the directions of maximum growth.

\bibliography{ref-LLT.bib}

\end{document}